\documentclass[conference, a4paper]{IEEEtran}
\IEEEoverridecommandlockouts

\ifCLASSINFOpdf
  \usepackage[pdftex]{graphicx}
  \DeclareGraphicsExtensions{.pdf,.jpeg,.png}
\else
\fi

\usepackage{amsmath,amssymb,amsfonts}
\usepackage{algorithmic}
\usepackage{graphicx}
\usepackage{textcomp}
\usepackage[left=1.4cm, right=1.4cm, top=1.7cm]{geometry}
\usepackage{caption}
\usepackage{anyfontsize}
\usepackage{xcolor}
\usepackage{array}
\usepackage{tabularx}
 \usepackage[natbib=true, style=numeric,sorting=none]{biblatex}
\usepackage{etoolbox}

\renewbibmacro*{doi+eprint+url}{%
  \iftoggle{bbx:doi}
    {\printfield{doi}\newunit\newblock} % <<-- add \newblock here for a line break
    {}%
  \iftoggle{bbx:eprint}
    {\printfield{eprint}\newunit\newblock}
    {}%
  \iftoggle{bbx:url}
    {\printfield{url}}
    {}}

\makeatletter
\newcommand*\titleheader[1]{\gdef\@titleheader{#1}}
\AtBeginDocument{%
\let\st@red@title\@title
\def\@title{%
\bgroup\normalfont\normalsize\centering\@titleheader\par\egroup
\vskip0.2em\st@red@title}
}
\makeatother

\makeatletter
\renewcommand{\fnum@figure}{Figure \thefigure}
\makeatother

\title{{Adaptive Traffic-Following Scheme \\ for Orderly Distributed Control of\\ Multi-Vehicle Systems} \\

\thanks{This work was funded by NASA's Convergent Aeronautics Solutions and Transformative Tools and Technologies projects.}
\vspace{0.5cm}
}

\titleheader{First US-Europe Air Transportation Research and Development Symposium (ATRDS 2025)}

\makeatletter
\newcommand{\linebreakand}{%
  \end{@IEEEauthorhalign}
  \hfill\mbox{}\par
  \mbox{}\hfill\begin{@IEEEauthorhalign}
}
\makeatother

\author{\IEEEauthorblockN{Anahita Jain}
\IEEEauthorblockA{The University of Texas at Austin\\
Austin, Texas \\
anaj18@utexas.edu
}
\and
\IEEEauthorblockN{Husni Idris}
\IEEEauthorblockA{NASA Ames Research Center\\
Moffett Field, California\\
husni.r.idris@nasa.gov}
\linebreakand
\IEEEauthorblockN{John-Paul Clarke}
\IEEEauthorblockA{The University of Texas at Austin\\
Austin, Texas \\
johnpaul@utexas.edu}
\and
\IEEEauthorblockN{Daniel Delahaye}
\IEEEauthorblockA{\'Ecole Nationale de l'Aviation Civile\\
Toulouse, France\\
daniel@recherche.enac.fr}
}

\renewcommand\IEEEkeywordsname{Keywords}
\IEEEaftertitletext{\vspace{-1\baselineskip}}

\begin{document}

\maketitle

\noindent \begin{abstract}

We present an adaptive control scheme to enable the emergence of order within distributed, autonomous multi-agent systems. Past studies showed that under high-density conditions, order generated from traffic-following behavior reduces travel times, while under low densities, choosing direct paths is more beneficial. In this paper, we leveraged those findings to allow aircraft to independently and dynamically adjust their degree of traffic-following behavior based on the current state of the airspace. This enables aircraft to follow other traffic only when beneficial. Quantitative analyses revealed that dynamic traffic-following behavior results in lower aircraft travel times at the cost of minimal levels of additional disorder to the airspace. The sensitivity of these benefits to temporal and spatial horizons was also investigated. Overall, this work highlights the benefits, and potential necessity, of incorporating self-organizing behavior in making distributed, autonomous multi-agent systems scalable.

\end{abstract}

\vspace{0.3cm}

\begin{IEEEkeywords}
disorder; order; entropy; autonomy; airspace operations; traffic pattern map; distributed; multi-agent
\end{IEEEkeywords}

\section{Introduction}

Autonomous vehicle operations are expected to increase in the airspace over the coming decades. Initially, applications will include non-passenger operations, such as fire fighting or cargo delivery using uncrewed aerial vehicles of different sizes. Eventually, the scope will expand to passenger-carrying vehicles for urban or regional air mobility. These vehicles are expected to interact and integrate with other traffic within the same airspace.

\begin{figure}[hbt!]
\centering
\includegraphics[width=0.48\textwidth]{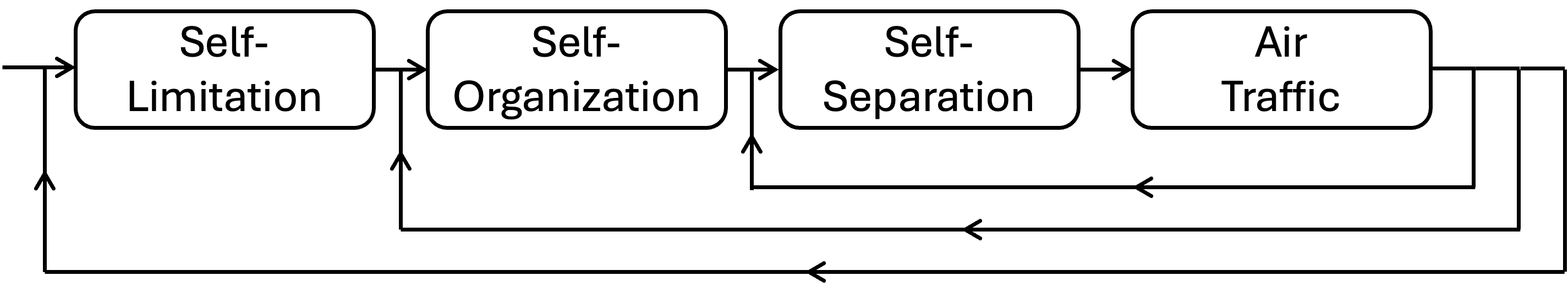}
\caption{Layered control loops of self-separation, self-organization and self-limitation.}
\label{fig: intro}
\end{figure}

For scalability, this airspace of the future will be a collective system of autonomous vehicles, where each vehicle makes increasingly independent decisions. Here, it is critical to design the system to allow the highest number of aircraft to use the airspace, while maintaining safety. To achieve this, a concept for digitally enabled vehicles was proposed with layered functions of self-separation in the inner loop and self-organizing and self-limiting behaviors in outer loops \cite{wing2022digital}, as shown in Fig. \ref{fig: intro}.

Self-organization is defined as the ability of a collective system of autonomous agents to create order in a distributed fashion. To maintain the orderliness of traffic flow in today's operations, air traffic controllers ensure aircraft compliance with established route structures and procedures; and apply equitable first-come, first-serve service. They also dynamically organize traffic into patterns to effectively mitigate complexity and workload challenges, especially in airspaces with high traffic densities \cite{b18}\cite{b17}. Therefore, creating order is important for scaling collective multi-vehicle autonomous systems to higher densities. 
% Therefore, incorporating self-organizing behavior within an airspace will enable it to maximize the number of aircraft that can navigate it safely.

% Self-limitation is defined as the ability of a collective system of agents to ensure separation distances are maintained, and agents have enough room to maneuver to mitigate the effects of uncertainties and perturbations. 

Self-limitation is defined as the ability of a collective system of autonomous agents to ensure sufficient maneuvering capability such that the inner-loop self-separation functions can maintain safe separation distances. In today's operations such limiting behavior is ensured by applying flow management techniques to limit flow rates or the number of aircraft in an airspace below acceptable controller workload limits.

We hypothesize that at low densities, self-separation- by aircraft maintaining minimum separation distances- is sufficient to ensure safety. As density increases, self-organization, and then self-limitation, become necessary to  maintain efficiency and safety. In this paper, we take a step towards enabling these functions by developing models and capabilities for a collective system of vehicles to dynamically create order in the self-organization loop. In the future, we will extend this study to the  self-limitation loop, and investigate how a system can limit itself to ensure safety.

Here, we assume a setting in which there is no central control authority enforcing strict structure and no controller workload constraints. We examine when maintaining order is necessary and beneficial, and when distributed vehicles should modify or relax such structures to improve their travel times. The dynamic emergence of order is particularly
useful in airspaces where structure does not exist \textit{a priori}, or is no longer applicable, for example, due to weather constraints. 
% This will become increasingly critical for enabling distributed multi-vehicle autonomous systems and their scalability to higher densities.

In addressing the need for orderly traffic behavior, researchers in the past have studied airspace order in the context of quantifying complexity from an air traffic controller's workload perspective such as in \cite{kopardekar2009airspace} or, more intrinsically, as in \cite{delahaye-entropy}. Order appears in convoy formation literature such as by maintaining close vehicle-to-vehicle alignment \cite{ishihara} or by following in the wake of a leader aircraft for energy saving benefits \cite{Xu}. Order was also observed as an emergent behavior from, for example, conflict resolution algorithms \cite{waltz2024self}\cite{con_res} and preserving vehicle trajectory flexibility for complexity mitigation mechanisms \cite{idris-delahaye}. 

% In addressing the need for orderly traffic behavior, the existing body of research has delved into airspace complexity metrics from an air traffic controller workload perspective such as in \cite{pk-dyndensity} or, more intrinsically, such as in \cite{delahaye-entropy}. Efforts have been made to study the emergence of orderly traffic patterns in distributed settings by utilizing flexibility metrics designed to mitigate high risk \cite{idris-delahaye} \cite{idrisaviation2011}. It has been shown that airspace complexity is reduced under such distributed schemes. Attempts have been made to mimic collective behavior from other social species such as ants for traffic management \cite{durant}. Convoy formation has also been studied to maintain vehicle-to-vehicle coherence and alignment for energy saving and benefits such as flow structuring \cite{ishihara} \cite{Xu}. 

In our previous publications \cite{jain2024benefits_jour}\cite{jain2024benefits}\cite{jain2024impact}, we investigated the dynamic emergence of traffic structures in a distributed multi-agent system. We developed a methodology to create a traffic pattern map of the airspace by leveraging information about the consistency and frequency of flow directions used by current as well as preceding traffic. Based on this map, cost functions were modeled to adapt to fixed levels of traffic-following behavior among aircraft within an airspace. Simulation results for this methodology showed that at low densities, traffic-following behavior resulted in a decrease in the entropy of the airspace with low penalties in terms of travel times. As the density of an airspace increased, substantial benefits in both airspace entropy and travel times were seen as the degree of traffic following behavior increased.

%\aj{one step closer to real-world}

This paper presents extensions to our previous work. So far, the degree of traffic-following behavior was set for all aircraft at the beginning of the simulation. This is akin to a central authority permanently assigning a fixed degree of traffic-following behavior for all aircraft within an airspace. In this paper, we propose a technique that allows aircraft to independently and dynamically update their degree of traffic-following behavior based on the density of traffic within their range. This has potential advantages because, as aforementioned, results from our previous analysis have showed that at low densities, low levels of traffic-following behavior are beneficial in terms of travel times whereas at high densities, increased levels of traffic-following behavior are beneficial. Therefore, by enabling aircraft to vary their degree of traffic-following behavior as they traverse the airspace, we enable them to adapt to dynamic airspace traffic volumes and patterns. Ultimately, this results in vehicles adopting higher degrees of traffic-following behavior at higher densities to maximize system throughput.

The rest of this paper is organized as follows: Section \ref{sec:method} contains distinct subsections that describe the modeling framework, path planning algorithm, quantitative metrics for measuring order, and the adaptive traffic-following factor. In  Section \ref{sec:results}, we present the experimental simulation setup and results exploring the effects of three factors on aircraft travel time and airspace entropy: (1) time-based discounting of the traffic pattern map, (2) fixed versus varying degrees of traffic-following behavior, and (3) varying spatial ranges in which traffic following is applied. In Section \ref{sec:future} we discuss potential applications of this work and in Section \ref{sec:future} we present conclusions and future research directions. 

\section{Methodology} \label{sec:method}

To investigate collective autonomous behavior that aims to maintain orderly traffic in a distributed manner, we employ models, algorithms, and metrics from \cite{jain2024benefits} and \cite{jain2024impact} with  extensions. These are divided into the following subsections: 
\ref{subsec:map}. A map of the airspace that depicts traffic patterns based on information from preceding traffic. We assume that this information is available to the vehicles either through their own sensors covering the airspace region of interest or through a service that gathers the information and broadcasts it to all vehicles. 
\ref{subsec:cost}. A cost function model that utilizes information from the traffic pattern map and calculates the amount of traffic-following behavior to apply relative to other utilities.
\ref{subsec:strategy}. A path planning algorithm that minimizes the cost function.
\ref{subsec:entropy}. A metric that measures order within an airspace.
\ref{subsec:traffic-following}. A technique that enables an aircraft to adjust its traffic-following behavior over the course of its flight.

% 1. A map of the airspace that depicts traffic patterns based on information from preceding traffic in Subsection \ref{subsec:map}. We assume that this information is available to the vehicles either through their own sensors covering the airspace region of interest or through a service that gathers the information and broadcasts it to all vehicles. 
% 2. A cost function model that utilizes information from the traffic pattern map and calculates the amount of traffic-following behavior to apply relative to other utilities in Subsection \ref{subsec:cost}.
% 3. A path planning algorithm that minimizes the cost function in Subsection \ref{subsec:strategy}.
% 4. A metric that measures order within an airspace in Subsection \ref{subsec:entropy}.
% 5. A technique to allow an aircraft to adjust its degree of traffic-following factor behavior over the course of its flight in Subsection \ref{subsec:traffic-following}.

\subsection{Traffic pattern map}\label{subsec:map}

\begin{figure}[hbt!]
\centering
\includegraphics[width=0.45\textwidth]{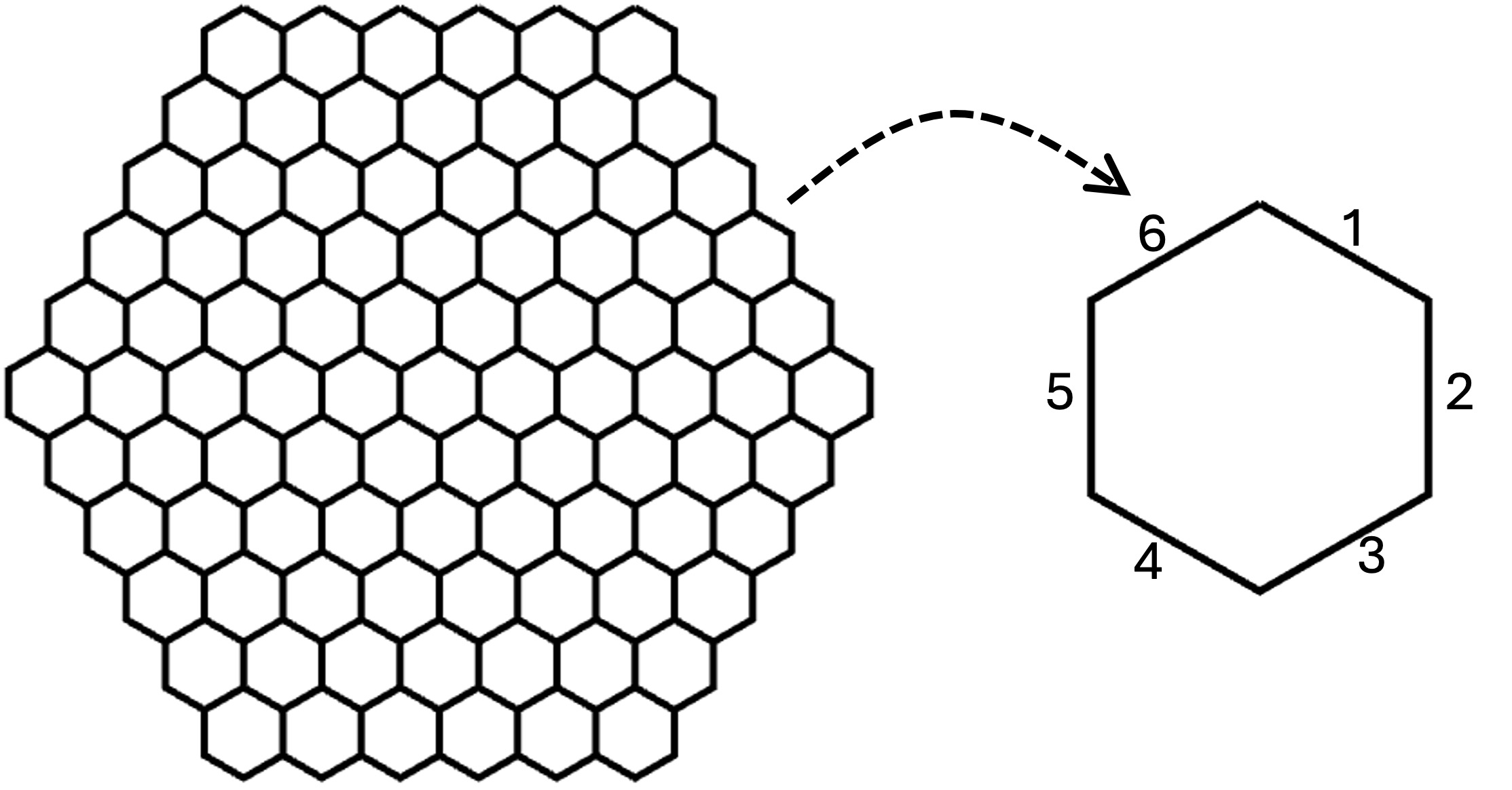}
\caption{
Partitioning the airspace into a hexagonal grid, with each cell assigned edges numbered 1 to 6.}
\label{hex}
\end{figure}

We assume the airspace is two-dimensional and partition it into regular hexagonal cells, tiled and pairwise congruent \cite{Patel_2013}, as depicted in Fig. \ref{hex}. Because our objective is to capture traffic directions, a hexagonal grid allows us to track more directions than, for example, a square grid, at reasonable computation costs. The edges of each cell are numbered as shown. 

For each possible flight traversal through the cell, we specify a corresponding ordered pair of edges in the format (entry edge $i$, exit edge $j$). Such a pair will be referred to as an \textit{edge pair}. Thus, for each hexagonal cell, a total of 36 edge pairs is possible. The number of traversals through an edge pair $(i, j)$ within a cell will be denoted by $t_{i, j}$. Thus, for each cell, a 6$\times$6 matrix denoted by Eq. \ref{t_matrix} records the number of aircraft that have traversed an edge pair of the cell so far, where the diagonal entries represent U-turns.

\begin{equation}
\label{t_matrix}
T = 
\begin{bmatrix}
    t_{1,1} & t_{1,2} & t_{1,3}& t_{1,4}& t_{1,5}& t_{1,6}\\
    t_{2,1} & t_{2,2} & t_{2,3}& t_{2,4}& t_{2,5}& t_{2,6}\\
    t_{3,1} & t_{3,2} & t_{3,3}& t_{3,4}& t_{3,5}& t_{3,6}\\
    t_{4,1} & t_{4,2} & t_{4,3}& t_{4,4}& t_{4,5}& t_{4,6}\\
    t_{5,1} & t_{5,2} & t_{5,3}& t_{5,4}& t_{5,5}& t_{5,6}\\
    t_{6,1} & t_{6,2} & t_{6,3}& t_{6,4}& t_{6,5}& t_{6,6}\\
\end{bmatrix}
\end{equation}

For example, if a cell has had no aircraft traversals in the past, the traffic matrix, T, is simply a 6$\times$6 zero matrix. Then, by time $\Delta $ if an aircraft has entered via the second edge and exited via the fifth and another aircraft has entered via the fourth and exited via the first, the corresponding T matrix of the cell is updated to: 
\begin{center}
$ T = \begin{bmatrix}
    0 & 0 & 0 & 0 & 0 & 0\\
    0 & 0 & 0 & 0 & 1 & 0\\
    0 & 0 & 0 & 0 & 0 & 0\\
    1 & 0 & 0 & 0 & 0 & 0\\
    0 & 0 & 0 & 0 & 0 & 0\\
    0 & 0 & 0 & 0 & 0 & 0\\
    \end{bmatrix}
$
\end{center}

% \begin{gather*}
%     T_{i, j} = 
%     \begin{cases} 
% 1, \text{ if } (i, j) = (2, 5) \text{ or } (4, 1)\\
% 0, \text{ otherwise}
% \end{cases}
% \end{gather*}

It is important to note here that non-zero entries signify the passage of traffic through the corresponding entry-exit pair, with the value indicating the number of aircraft that have used the same pathway. By using this technique, we capture not only the different directions of traffic through a cell, but also the number of aircraft that have traversed the cell in these directions. As these entries accumulate over time, we create a map of all traffic patterns in the airspace.

\subsection{Cost function model} \label{subsec:cost}

In the distributed environment setup used in this paper, any aircraft equipped to use this algorithm, called ``ownship," makes its own path-planning decision using a cost function. An ownship incurs a cost whenever it passes through a cell from any one edge to another. We divide these costs into two parts: \textit{unimpeded transit costs} and \textit{traffic costs}, which are detailed in the following paragraphs.

\subsubsection*{Unimpeded transit cost through a cell}
The cost of unimpeded transit through an edge pair  $(i, j)$ is denoted by $u_{i, j}$. This value can be adjusted based on how expensive it is for an ownship to transit an edge pair based on winds, weather, or other non-traffic related factors within the airspace. For example, edge pairs encompassing bad weather can be given higher costs of transit to deter aircraft from traveling through the airspace they encompass. Therefore, less favorable transit pairs may be set up to have higher costs and vice-versa. 

For each hexagonal cell, there are 36 unique values denoting transit cost between each edge pair which are conveniently stored in a 6 $\times$ 6 matrix as follows.

\begin{equation}
    \label{u_matrix}
    U = 
\begin{bmatrix}
    u_{1,1} & u_{1,2} & u_{1,3}& u_{1,4}& u_{1,5}& u_{1,6}\\
    u_{2,1} & u_{2,2} & u_{2,3}& u_{2,4}& u_{2,5}& u_{2,6}\\
    u_{3,1} & u_{3,2} & u_{3,3}& u_{3,4}& u_{3,5}& u_{3,6}\\
    u_{4,1} & u_{4,2} & u_{4,3}& u_{4,4}& u_{4,5}& u_{4,6}\\
    u_{5,1} & u_{5,2} & u_{5,3}& u_{5,4}& u_{5,5}& u_{5,6}\\
    u_{6,1} & u_{6,2} & u_{6,3}& u_{6,4}& u_{6,5}& u_{6,6}\\
\end{bmatrix}
\end{equation}
\subsubsection*{Traffic cost through a cell}

There are a number of ways that traffic can affect an ownship. In order to maintain safe minimum separation requirements among aircraft, the number of aircraft is limited to one in each cell. In this cost function, we model the inclination of an aircraft to follow current and past traffic patterns within an airspace as follows. 

For each edge pair, we define a traversal cost based on the number of aircraft that have used the pair. This cost is formulated as $(1 - k_t*{\hat{t}_{i, j}})$ where $k_t$ is the \textit{traffic-following factor} and $\hat{t}_{i, j}$ is the corresponding normalized entry from Eq. \ref{t_matrix}. Subtracting the normalized traffic count from 1 makes the edge pairs with more traffic in them less costly to the ownship, thus making pairs with higher traffic ``attractive" to the ownship. The traffic-following factor, $k_t$ is adopted as a gain in order to tailor the degree to which traffic is attractive to an ownship. The higher the value of $k_t$, the less costly it becomes for an ownship to use the edge pair, i.e., the more inclined an ownship is to follow traffic. A technique to update $k_t$ for an ownship throughout the course of its flight is discussed in detail in Section \ref{subsec:traffic-following}.

Similar to previous setups, for each cell, there are 36 such values, which are stored in a 6$\times$6 matrix. Hence, the traffic cost through different entry-exit pairs in a cell is: 

\begin{equation}
    \label{t_cost}
    \left[
\textbf{1}_{6x6} - k_t\frac{T}{\sum_{i, j}t_{i, j}}
\right]
\end{equation}

% \begin{equation}
%     \label{t_cost}
%     \left[
% \textbf{1}_{6x6} - k_t\frac{T}{\sum_{i, j}t_{i, j}}
% \right]
% \end{equation}

% where $\textbf{1}_{6x6}$ is a 6 $\times$ 6 matrix of ones and 

where $\textbf{1}_{6x6}$ is a 6 $\times$ 6 matrix of ones and $\sum_{i, j}t_{i, j}$ is the grand sum of the traffic matrix of that cell. 

The current approach for quantifying traffic costs (or attractiveness) considers similarity in the direction of aircraft and vehicular count. In future work, we plan on incorporating additional characteristics such as similarity between speeds and sizes of different agents. 

\subsubsection*{Total cost of transit through a cell} 
To get the total cost of transit between an edge pair of a cell, we simply sum up the corresponding unimpeded transit and traffic costs. For each cell, these costs can be stored in a $6\times6$ matrix as follows. 

\begin{equation}
    \label{final_cost_equation}
    C = U + 
    \left[
\textbf{1}_{6x6} - k_t\frac{T}{\sum_{i, j}t_{i, j}}
\right]
\end{equation}

% \begin{equation}
%     \label{final_cost_equation}
%     C = U
%  +  \left[
% \textbf{1}_{6x6} - k_t\frac{T}{\sum_{i, j}t_{i, j}}
% \right]
% \end{equation}

\subsection{Path planning} \label{subsec:strategy}
Here, an algorithm that uses the cost function defined in Section \ref{subsec:cost} to plan an ownship's trajectory is covered.

\subsubsection*{Least costly path calculation}

For every possible path between the initial and final position, we compute a cost using the costs introduced in Eq. \ref{final_cost_equation}. 

We cast the path-planning problem by introducing an undirected graph \cite{b16}, whose nodes are all the edges of the cells\footnote{In the graph theoretic approach used here, the terms nodes and arcs are used instead of the more commonly used terms vertices and edges, to avoid confusion with the hexagonal grid's geometry.}. There are only two sets of possible arcs defined in this setup: 1. Two nodes corresponding to two edges of the same cell are connected by an arc that equals the total cost of traversal of that edge pair. 2. Two nodes corresponding to the coincident edges of neighboring cell are connected by an arc that equals the cost of 0.

An ownship computes its trajectory using this weighted graph setup. To calculate the least costly path, we now use Dijkstra's algorithm \cite{b9}. Starting from the initial input edge, the algorithm traverses the hexagonal grid moving from one edge of a cell to all other edges in the cell, finding the least costly path. To dynamically account for updated costs due to changes in traffic, an ownship re-calculates its least costly path to its destination from its current position periodically. This enables it to choose the least costly path throughout the course of its trajectory. For all simulations covered in this paper, an ownship recalculated its best path forward every time it entered a new cell.

\subsubsection*{Updates to Ownship State}
Every time an ownship enters a new cell, it determines the next best edge to travel to by using the path planning algorithm from its current position in the grid to its destination. 
In the experiments here, we omit the addition of the conflict resolution algorithm \cite{con_res} as done in \cite{jain2024impact}. Instead, we limit cell capacity to one, i.e, allow the presence of no more than one vehicle within a cell at any given time. We assume this capacity limit, along with the cell size, ensures aircraft do not violate minimum horizontal separation requirements. 

% \aj{assuming that separation is not violated; sufficient room is }
% At the same time, it eliminates any possible emergent behavior caused by the conflict resolution algorithm \aj{great to cite something here}. 

If an ownship's next best cell is occupied, it holds its current cell for a maximum additional time period, $t_{\text{hold}}$. During this period, if the next best cell empties, the ownship proceeds to enter it. If it is unsuccessful in moving forward and has been waiting for the cell to be empty for $t_{\text{hold}}$ seconds, it will recalculate its best path forward to find the next best cell to move towards. Once an ownship enters a new cell, the traffic pattern map records its traversal through the corresponding edges of the last cell.

Updates to an ownship's trajectory are made in the form of a cumulative heading change, $\Delta\psi$, while speed and altitude are kept constant. In a simulation with multiple ownships, each ownship's trajectory is updated sequentially (with the traffic pattern map available when needed), with no inter-agent coordination or negotiation, at every time-step.

% \aj{talk about updating $k_t$ for each ownship}

\subsection{Entropy}\label{subsec:entropy}
This section details how we define order within the context of our work. To measure the level of order in our system, we use a popular formula used to measure entropy in the field of information theory \cite{b10}. For a discrete random variable $\textit{X}$ that is distributed according to ${\displaystyle p\colon {\mathcal {X}}\to [0,1]}$, 
the entropy is

\begin{equation}
    \label{shannon}
    {\displaystyle \mathrm {H} (X)=-\sum _{x\in {\mathcal {X}}}p(x)\log p(x)}
\end{equation}

where $x$ is a value from set $\textit{X}$, $p(x)$ is the probability of $x$ occurring in $\textit{X}$, and $\sum$ denotes the sum over the variable's possible values. 

It is important to note here that Eq. \ref{shannon} can be used for measuring the entropy of various factors in an airspace. Options include metrics such as the magnitude of total heading changes within an airspace, transit time for an aircraft, etc. 

For the scope of this analysis, consider the airspace configuration composed of hexagonal cells. For a single cell, we quantify entropy based on two factors: (1) the diversity of directions (entry-exit pairings) through which traffic flows; and (2) the number of aircraft using these pathways. As stated earlier, the traffic matrix, T, stores information about the number and directions of crossings between each entry-exit pairing in a cell over time. Therefore by leveraging the traffic matrix T associated with a cell as the variable in Eq. \ref{shannon}, we calculate the entropy within that individual cell. Subsequently, by summing entropy values across all cells within a grid, we obtain a metric that encapsulates the total entropy of the airspace up until that time.

Detailed examples explaining the use of Eq. \ref{shannon} in the context of this work can be found in \cite{jain2024benefits}. A higher entropy value signifies lower order, greater disorder, or simply a higher diversity of pathways used within a system.

\subsection{The traffic-following factor, $k_t$} \label{subsec:traffic-following}

\begin{figure}[hbt!]
\centering
\includegraphics[width=0.5\textwidth]{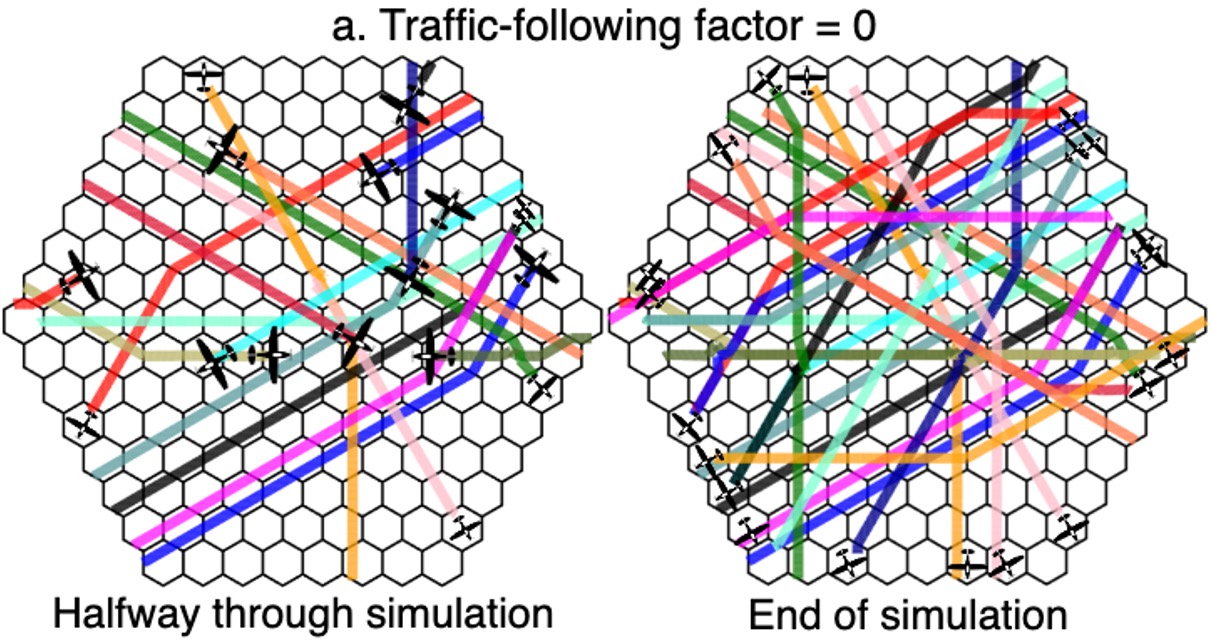}
\includegraphics[width=0.5\textwidth]{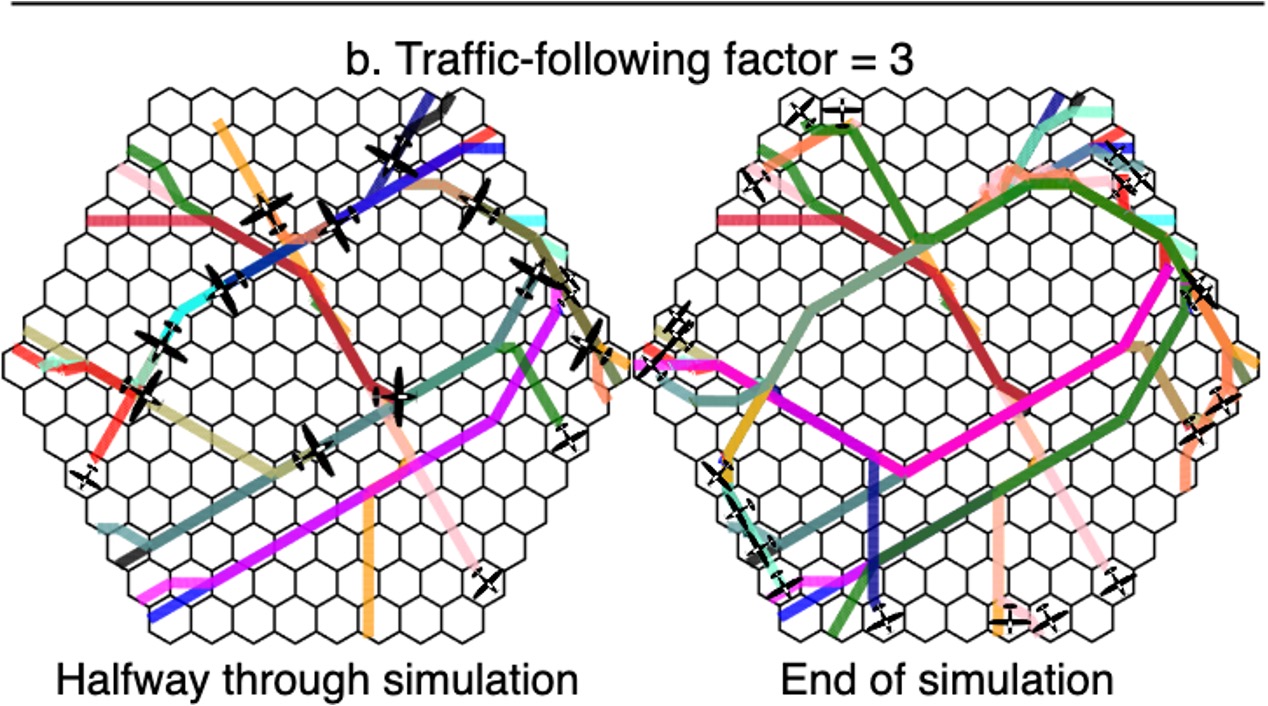}
\caption{Simulation snapshots demonstrating the effect of $k_t$ on traffic patterns when all aircraft are ownships, showing that order emerges when aircraft follow patterns.} 
\label{fig:results}
\end{figure}

As discussed in Section \ref{subsec:cost}, the traffic-following factor $k_t$, is a gain that determines the degree to which an ownship is attracted to traffic. A high value of $k_t$ means it is less costly for an ownship to use a path (an entry-exit edge pair) being used by other traffic, i.e., the ownship is more inclined to follow preexisting flight patterns, as seen in Fig. \ref{fig:results} \cite{jain2024impact}. 

\begin{figure}[hbt!]
\centering
\includegraphics[width=0.5\textwidth]{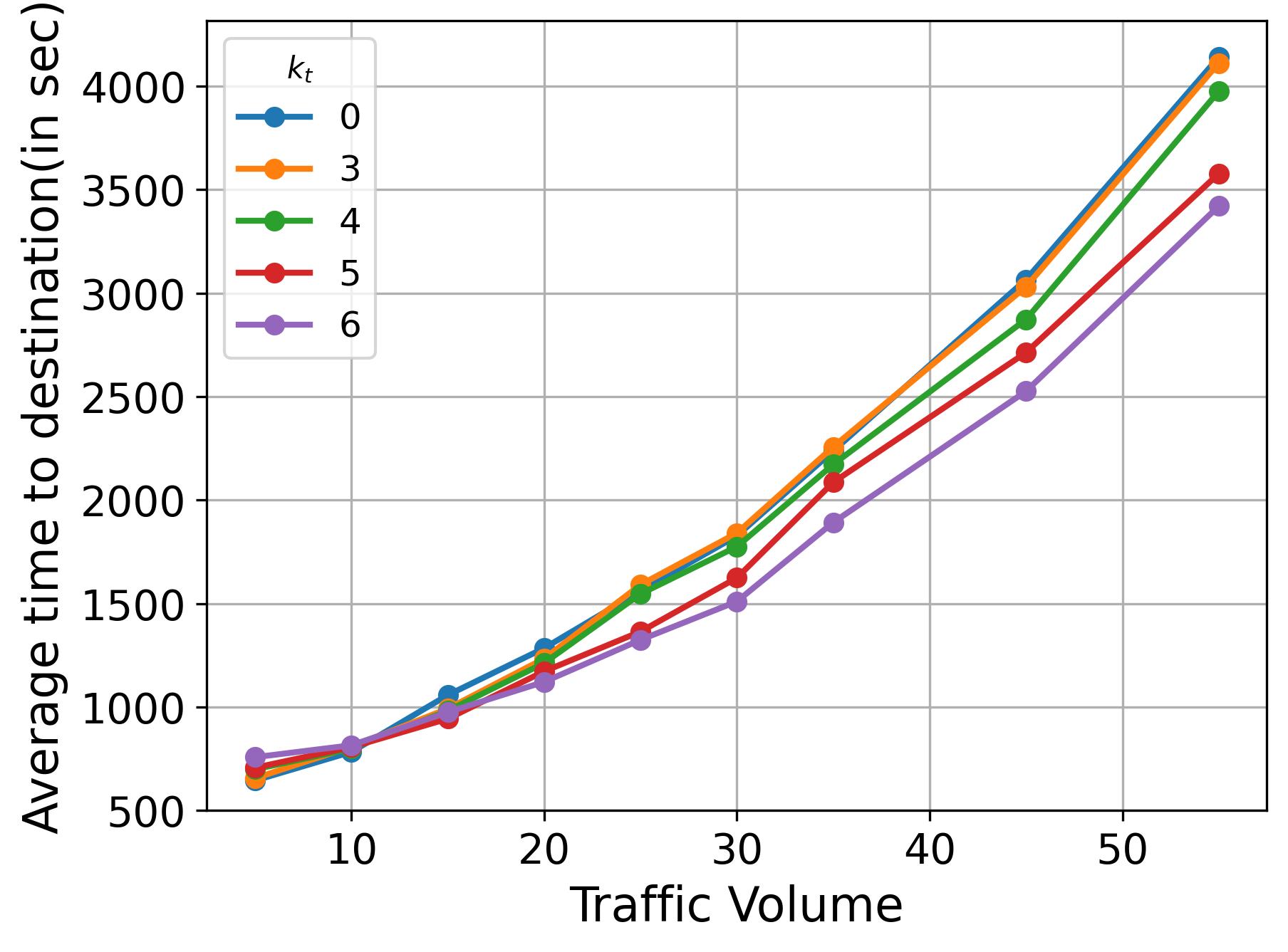}
\caption{Results showing the effect of fixed values of the traffic-following factor ($k_t$) on average aircraft travel times for different traffic volumes from \cite{jain2024benefits}. } 
\label{init_results}
\end{figure}

Previous studies on this subject, \cite{jain2024benefits} \cite{jain2024impact}, fixed $k_t$ to a constant value for all aircraft in the airspace throughout the course of the simulation. Studies conducted so far ran the same set of aircraft (same origin-destination pairs) for multiple runs, wherein the only factor varying between runs was the preset fixed value of the degree of traffic-following behavior that aircraft exhibited. This technique was adopted to analyze the impact of traffic-following behavior on travel times and airspace order under varying traffic density levels. Fig. \ref{init_results} shows the effects of traffic-following behavior on aircraft travel times from \cite{jain2024benefits}. Results showed that, at low densities, traffic-following behavior resulted in a decrease in the entropy of the airspace with low penalties in terms of travel times. As the density of the airspace increased, substantial gains in both airspace entropy and travel times were seen as the degree of traffic following behavior increased.

In this extension of the work, we leverage data from past studies to enable aircraft to gracefully adjust their degree of traffic-following behavior independently, throughout the course of their flight, to dynamically account for changing airspace conditions. By referring to past results, as shown in Fig. \ref{init_results}, we model a function that will output the ideal value of the traffic-following factor $k_t$, given the amount of vehicular congestion detected by an ownship. This ideal value of $k_t$ will enable an aircraft to reach its destination in the shortest possible time, given the current state of the airspace. 

We begin by defining a range $R_s$ for each ownship. This value indicates the radius of a circle (sphere if the airspace were three-dimensional) that represents sensory limitations, or may simply be a way to limit an ownship's traffic-following behavior to a neighborhood around it. We define $\rho$ as the density of aircraft in an ownship's range $R_s$. This is calculated by dividing the number of aircraft in an ownship's range $R_s$, by the area of the range. Data collected from past studies is based on aircraft being able to have perfect information about each other's position, regardless of inter-agent distance. In essence, each agent's range was set to the entire grid. Hence, to use data from Fig. \ref{init_results}, we must first adjust for range $R_s$.
% as shown in Table \ref{tab:1}.

% \begin{table}
% \begin{center}
% \begin{tabularx}{0.3\textwidth}{ 
%   | >{\centering\arraybackslash}X 
%   | >{\centering\arraybackslash}X 
%   | >{\centering\arraybackslash}X | }
%  % \caption{A Table}\label{tab:1}\\
%  \hline
%  $k_t$ & N & $\rho$ \\
%  \hline
%  0 & 1  & 0.001  \\
%  0 & 5  & 0.003  \\
%  0 & 10  & 0.006  \\
%  5 & 15  & 0.008  \\
%  6 & 20  & 0.011  \\
%  6 & 25  & 0.014  \\
%  6 & 35  & 0.020  \\
%  6 & 45  & 0.025  \\
%  6 & 55  & 0.031  \\
% \hline
% \end{tabularx}
% \end{center}
% \caption{Past data points indicating which traffic-following factor ($k_t$) values yield lowest travel times at specific airspace traffic volumes (N) were leveraged. By adjusting these values for airspace density ($\rho$), we are able to find a function to adjust $k_t$ for a given traffic density.}
% \label{tab:1}
% \end{table}

% Columns 1 and 2 in Table \ref{tab:1} are data points from Fig. \ref{init_results} corresponding to the least travel time. Since values contained in Column 2 are for the entire grid, Column 3 shows values adjusted for density, i.e., the number of aircraft from Column 2 divided by the total area of the grid (1786.78 square miles). It is interesting to note that the crossover point from the minimum ($k_t$ = 0) to maximum ($k_t$ = 6) occurs quickly, suggesting that sigmoid curve might be sufficient for a continuous traffic-following function. 

We select data points from Fig. \ref{init_results} corresponding to the least travel time. Since traffic volume numbers are for the entire grid, we adjust these values for density, i.e., the traffic volume value is divided by the total area of the grid (1786.78 square miles). It is interesting to note that the crossover point from the minimum ($k_t$ = 0) to maximum ($k_t$ = 6) occurs quickly, suggesting that a sigmoid curve might be sufficient for the continuous traffic-following function.

\begin{figure}[hbt!]
\centering
\includegraphics[width=0.4\textwidth]{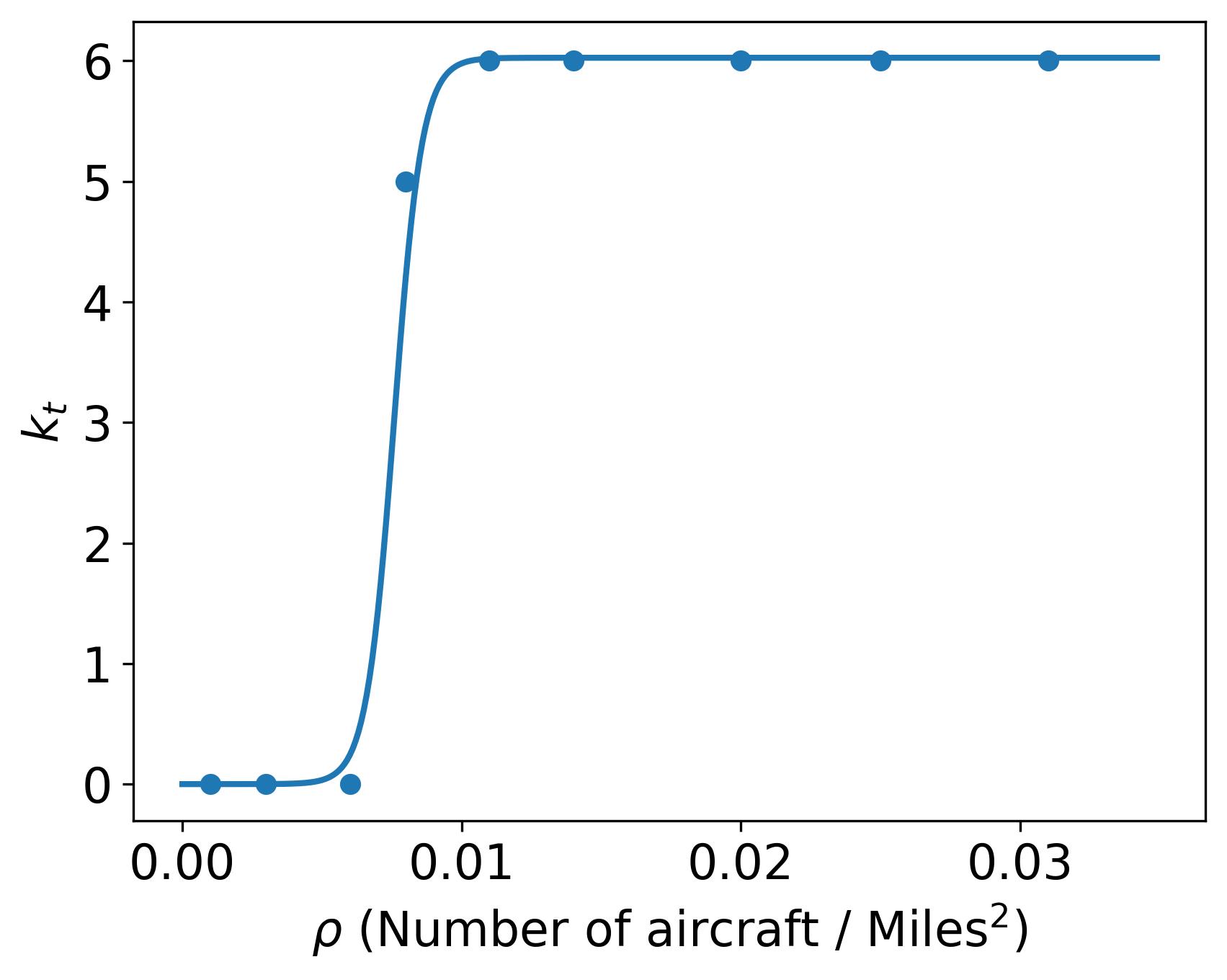}
\caption{Sigmoid function from Eq. \ref{eq:k_t} illustrating how the traffic-following factor ($k_t$) varies with the density of aircraft in an ownship's range ($\rho$).} 
\label{fig:k_t}
\end{figure}

Therefore, these values were then fit into a sigmoidal function (as shown in Fig. \ref{fig:k_t}) to get the following equation:

\begin{equation}
\label{eq:k_t}
    % Y_fit = y_max/(1+e^(-(x/slope - int)))
    k_t = \frac{6.024}{1+e^{(-\frac{\rho}{0.0005} - 15.193)}}
\end{equation}

where $\rho$ is the density of traffic in an ownship's range $R_s$ with units number of aircraft per square mile. Thus, an ownship is able to determine the ideal degree of traffic-following behavior based on the amount of traffic in its range.

\section{Results} \label{sec:results}
We begin this section by describing the experimental setup for the simulations conducted in this study and presenting an example that illustrates how the traffic-following factor was adjusted for varying levels of traffic congestion. Then, Section \ref{subsec:disc} covers the effects of discounting the traffic pattern map with time. Section \ref{subsec:const_vs_varying} covers the effects of fixed versus dynamically changing the degree of traffic-following behavior on travel times and airspace entropy. Section \ref{subsec:varying_ranges} covers how range $R_s$ impacts travel times and airspace entropy when aircraft dynamically update their traffic-following behavior.

\subsection*{Experimental setup}

For all simulations conducted, all aircraft were treated as ownships traveling at a constant speed of 250 miles per hour. Each ownship was assigned randomly generated initial and final positions. Within the grid, these position coordinates were mid-points of any outward-facing edges of the outermost ring of cells. To increase the path-length traversed by the ownships, and thus increase inter-agent interaction, initial and final positions were assigned such that they did not lie on adjacent edges of the outermost ring of hexes. 

% \aj{speed}
The grid was assumed to be two-dimensional. Each hexagonal cell had an edge length of 2.5 miles, which is also the distance from the center of the cell to any of its vertices. For each cell, the cost of unimpeded transit $u_{i, j}$, through an edge pair $(i, j)$ was set to be the distance between the edge midpoints, without any adjustments for wind and weather. For U-turns, the cost was approximated as four times the edge-length of a cell. Here, $t_\text{hold}$ was set to 150 seconds.

In all experiments studied in the following subsections, $N$ = \{4, 5, 4, 5, 40, 20, 10, 4, 5, 6, 4, 3, 10, 40, 20, 10, 5, 10, 30, 20, 10, 6, 5, 3$\}$ aircraft were introduced into the airspace at \{0, 500, 1000, 1500, 2000, 2500, 3000, 3500, 4000, 4500, 5000, 5500, 6000, 6500, 7000, 7500, 8000, 8500, 9000, 9500, 10000, 10500, 11000, 11500\} seconds from the start of the simulation. Such a traffic profile was chosen to ensure testing at varying congestion levels. For each study, we ran 15 simulations (found sufficient by statistical analysis), each with a different set of randomly generated initial and final coordinates for each aircraft.

In Section \ref{subsec:disc}, we summarize the effects of discounting the traffic pattern map with time. Based on results from Section \ref{subsec:disc}, we incorporated a time-based discounting of the traffic pattern map in Sections \ref{subsec:const_vs_varying} and \ref{subsec:varying_ranges}. In these two studies, ownship based their decision-making on traffic patterns from the preceding 500 seconds at any given moment. For cases in which the traffic-following factor was dynamically adjusted by an ownship, each ownship adjusted its traffic-following factor every 100 seconds using Eq. \ref{eq:k_t}.

% It should be noted that results shown below might vary based on the traffic profile used. 
% For instance, a traffic profile with only low volumes of traffic might 

\begin{figure}[hbt!]
\centering
\includegraphics[width=0.45\textwidth]{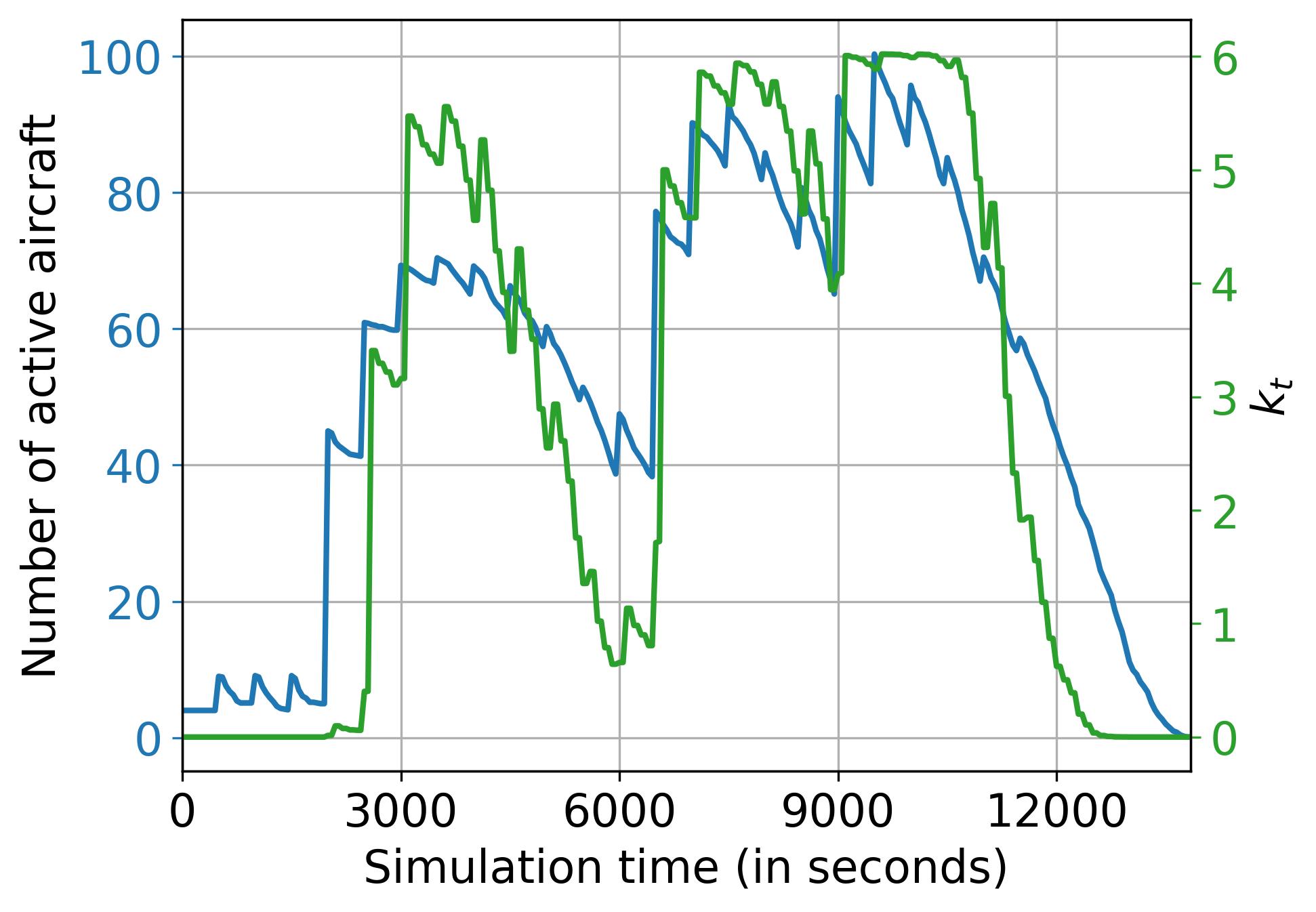}
\caption{Plot depicting an ownship's adjustment of the traffic-following factor $k_t$ (in green), in response to the density of aircraft within its range $R_s$ (in blue), where $R_s$ encompasses the entire grid in this example.} 
\label{k_t vs num}
\end{figure}

For the traffic profile used in all subsections, Fig. \ref{k_t vs num} shows an example for the case where $k_t$ was adjusted for all aircraft according to Eq. \ref{eq:k_t}, with all aircraft having a range spanning the entire grid, $R_s$ = 50 miles. The X-axis represents time passed from the beginning of the simulation. The Y-axis on the left corresponds to the traffic profile used in the simulation (in blue) whereas the Y-axis on the right corresponds to how the traffic-following factor, $k_t$, changed throughout the course of the simulation (in green). The sigmoidal nature of the $k_t$ curve is also observed here, with values increasing and decreasing rapidly as density changes. As anticipated, the traffic-following factor remains low when there are fewer aircraft in the airspace but increases  as the number of aircraft in the airspace grows. We hypothesize this will benefit aircraft in terms of travel times, as past results have shown that increased traffic-following behavior at high densities leads to improvements in travel times while decreased traffic-following behavior is most beneficial at low densities.

% \subsection{The effect of discounting traffic information on travel time and entropy}\label{subsec:disc with time} 

\subsection{The effect of discounting the traffic-pattern map with time on travel time and airspace entropy} \label{subsec:disc}

The traffic pattern map discussed in Section \ref{subsec:map} accumulates information about the variety of directions aircraft use to navigate the airspace, along with the number of aircraft following these directions over time. In this study, we investigated how an ownship’s travel time changes based on the history of the traffic pattern map available to it.

We studied the traffic profile described in the experimental setup above under two conditions. In the first case, we incorporated a time-based discounting of the traffic pattern map, % without considering traffic patterns from time points preceding 500 seconds at any given moment. 
where traffic patterns that occurred prior to 500 seconds from the current simulation time were ignored. While in this study we chose to simply not consider data from time points earlier than a threshold, in general the discounting can also be modeled as an exponential time-decay function. In the second case, aircraft did not discount past traffic patterns at all. This meant that if an aircraft adjusted its traffic-following factor to a non-zero value in this case, the traffic patterns available to it were not temporally differentiated, i.e, they had accumulated since the beginning of the simulation. In both cases, all aircraft independently varied their degree of traffic-following behavior, and their range was set to the entire grid. Although the threshold of 500 seconds was found to be adequate for this experiment based on some trial runs, a formal study on the effect of the value of discounting threshold will be conducted in the next phase of this work.

\begin{figure}[hbt!]
\centering
\includegraphics[width=0.45\textwidth]{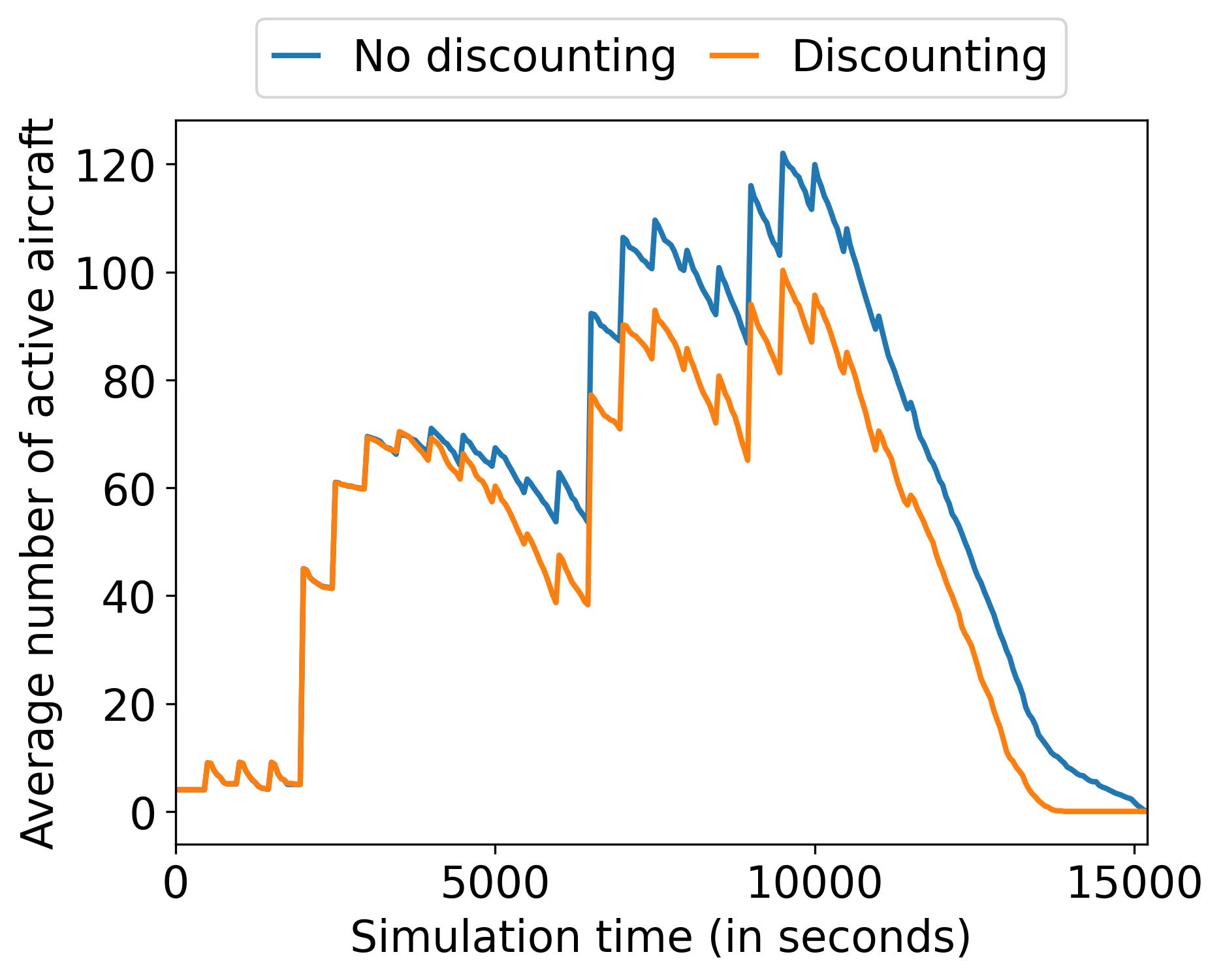}
\caption{Effect of discounting the traffic pattern map with time on the average number of aircraft present in the airspace.} 
\label{sec3_num}
\end{figure}

Fig. \ref{sec3_num} shows how the number of aircraft in the airspace changed over the course of the simulation in both cases, even though they were introduced at the same time in each case. While both cases have a similar number of aircraft in the system at the beginning of the simulation, the case with discounting has consistently fewer aircraft than the case without discounting towards the middle and end of the simulation. This indicates that airspace congestion was higher in the case of no discounting.

% The case in which the traffic pattern map is not discounted with time consistently shows higher levels of traffic congestion compared to when the traffic pattern map is discounted with time.

\begin{figure}[hbt!]
\centering
\includegraphics[width=0.45\textwidth]{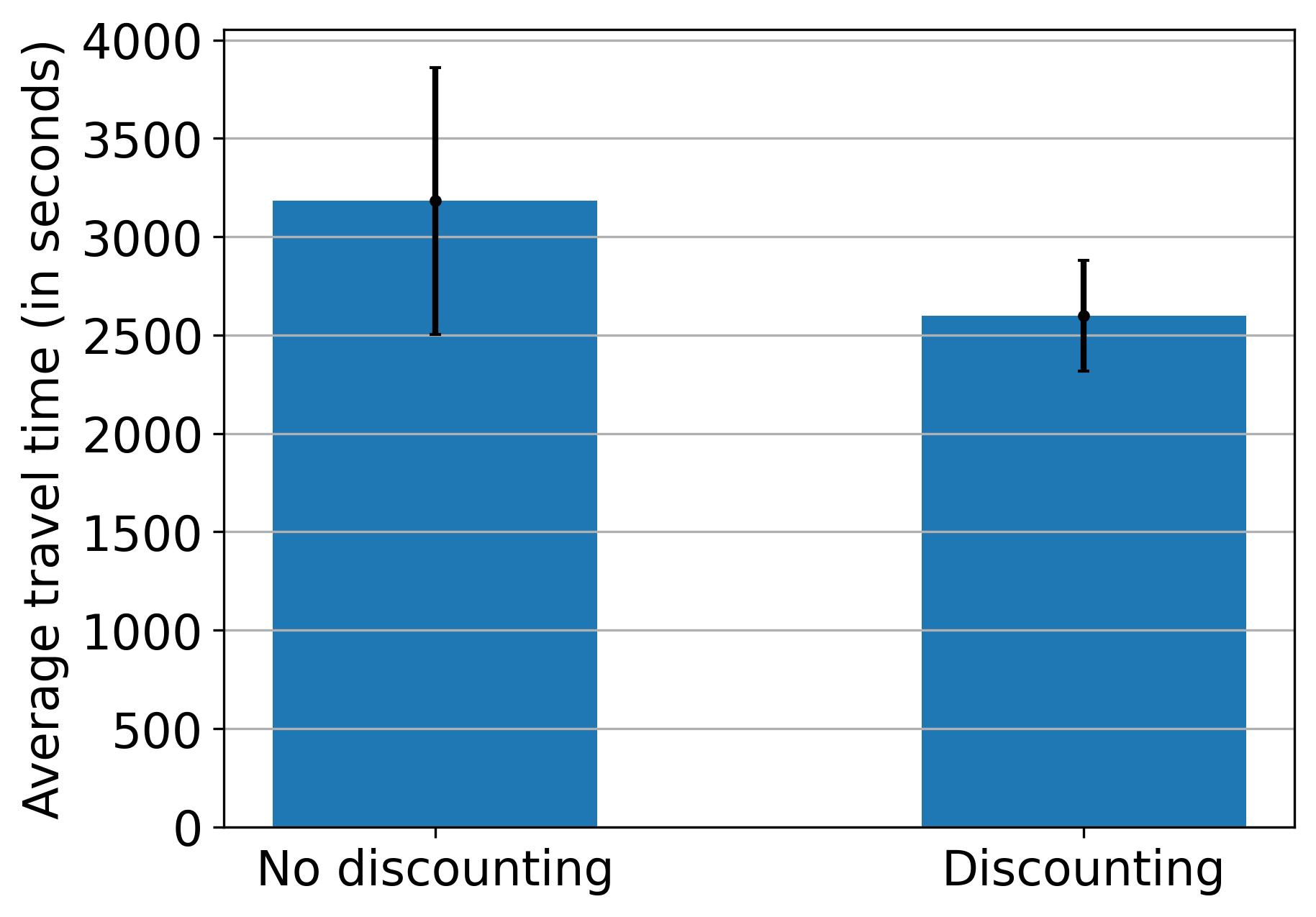}
\caption{Effect of discounting the traffic pattern map with time on average aircraft travel times.} 
\label{sec3_ttd}
\end{figure}

Results for the average amount of time it took an ownship to traverse the airspace in each case are shown in Fig. \ref{sec3_ttd}. An aircraft took 18\% less time on average (p-value = 2.19E-13) to reach its destination in the case when the traffic pattern map was discounted as compared to when it was not.

\begin{figure}[hbt!]
\centering
\includegraphics[width=0.45\textwidth]{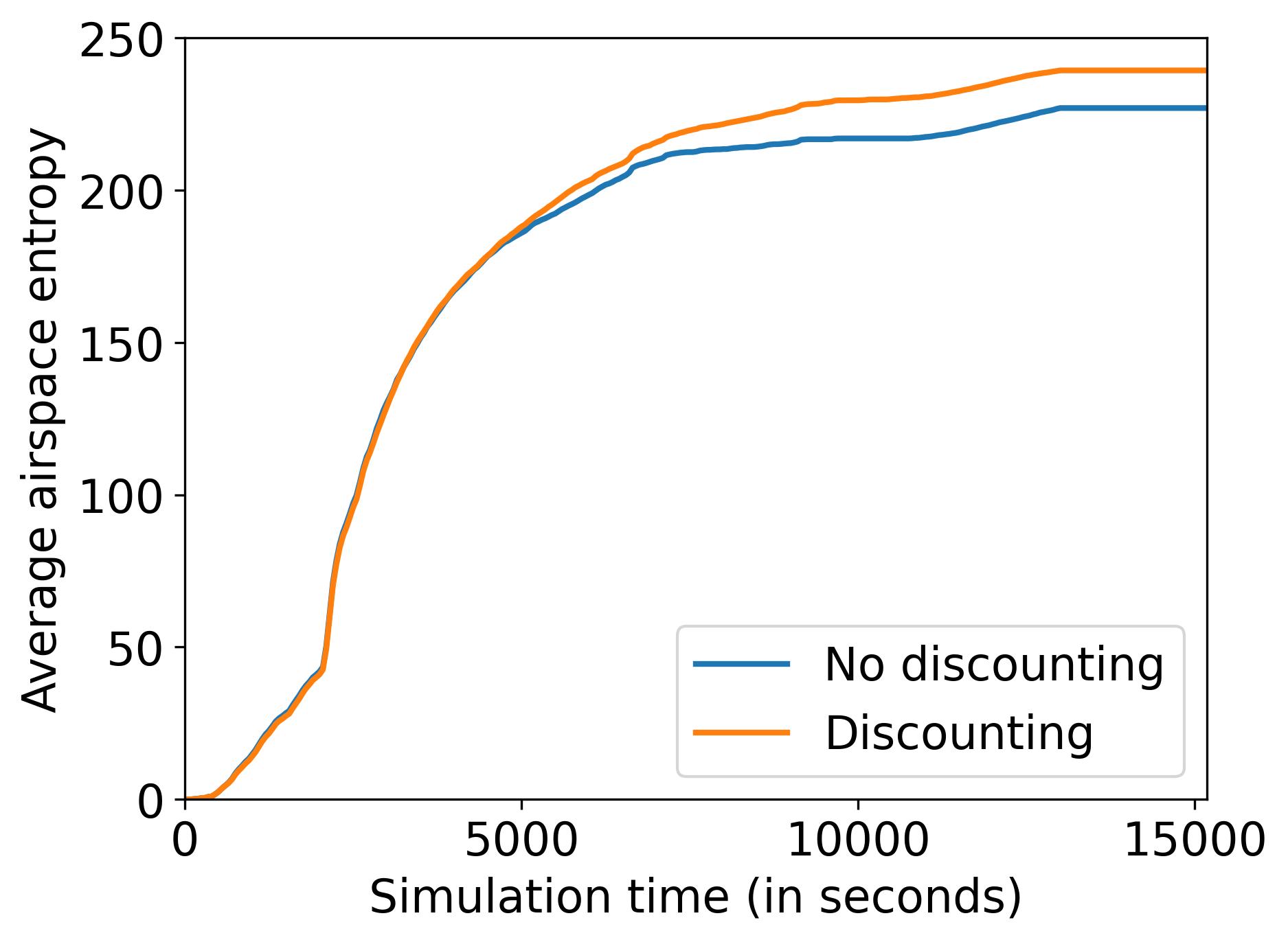}
\caption{Effect of discounting the traffic pattern map with time on airspace entropy.} 
\label{sec3_entropy}
\end{figure}

Results for the average airspace entropy over the course of the simulation for each case are shown in Fig. \ref{sec3_entropy}. In both cases, airspace entropy increased over time. This aligns with the principle that, in an isolated system, the total degree of chaos or disorder will either increase or remain constant, but never decrease. In this system, entropy is calculated based on the diversity of directions through which traffic flows and the number of aircraft that use these directions. Both these factors can either increase over time as more aircraft traverse the airspace, or stay constant if we see no new directions, but never decrease. This also explains why all scenarios showed a stagnation in entropy towards the end of the simulation. From Fig. \ref{sec3_entropy} we observe that the amount of disorder added to the airspace was consistently lower, albeit not by much, in the case when the traffic pattern map was not discounted.

% Trends observed from this analysis are logically consistent. 

To understand these trends, consider the case when the traffic pattern map is not discounted with time. Because at the beginning of the simulation traffic volumes in the airspace are low, aircraft adjust their paths to exhibit low degrees of traffic-following behavior. In doing so, aircraft move in ``selfish" patterns, utilizing more varied directions in the airspace. 

As time passes and the volume of traffic in the airspace increases, aircraft adjust their traffic-following factors to higher values. This can be potentially beneficial since it encourages more orderly traffic by leveraging flows collected in the traffic pattern map. However, in this case, the traffic pattern map holds aircraft trajectories amassed since the beginning of the simulation, i.e., it includes paths that were taken by more ``selfish" aircraft during low-density periods. Following these paths might not benefit aircraft in the high-density period since they also provide less beneficial (more travel time) options. Since the traffic pattern map is not discounted with time in this case, aircraft still follow these paths without the benefits of faster travel times.

Therefore, when older traffic patterns are not discounted, the creation of dynamic paths based on more recent airspace pattern trends is not encouraged, resulting in lower airspace entropy. At the same time, older, historic paths are used at the cost of increased travel times. On the other hand, if the traffic pattern map is discounted based on time, only more recent patterns are available to aircraft choosing to follow traffic, leading to lower travel times. Therefore, the introduction of discounting to the traffic pattern map makes traffic-following more robust to varying levels of congestion in the airspace.

% If the traffic pattern map is discounted based on time, only more recent patterns are available to aircraft choosing to follow traffic, leading to lower travel times. On the other hand, if there is no temporal discounting of the traffic pattern map, patterns simply accumulate over time. The addition of this time-based discounting in the traffic pattern map is beneficial when density transitions from low to high while it has no impact, negative or positive, when density changes from high to low. Therefore, the introduction of discounting to the traffic pattern map, makes this work more robust to varying traffic profiles. It should be noted that studies covered in \cite{jain2024benefits} and \cite{jain2024impact} were not affected by this accumulation of data in the traffic pattern map since they did consider varying traffic profiles.

Trends observed in Figs. \ref{sec3_num}, \ref{sec3_ttd}, and \ref{sec3_entropy} encouraged us to incorporate a time-based discounting of the traffic pattern map. Hence, in the rest of the experiments presented in this paper, we introduced a time-based discounting of the traffic pattern map considering only more recent traffic patterns.

\subsection{The effect of fixed versus varying degrees of traffic-following behavior on travel time and airspace entropy}\label{subsec:const_vs_varying}

This experiment examines how an ownship's ability to vary its degree of traffic-following behavior compares to scenarios when it is fixed at specific values. For this, simulations were run to compare four cases where the degree of traffic-following behavior was fixed ($k_t = \{0, 3, 5, 6 \}$) to a case where aircraft were allowed to independently vary their degree of traffic-following behavior. Since the cases in which $k_t$ is fixed have no range variation, the range $R_s$ was set to the entire grid (50 miles) for the varying $k_t$ case. It should be noted here that when $R_s$ spans the entire grid, each aircraft considers all aircraft being in its range, resulting in all aircraft selecting the same value for $k_t$.

\begin{figure}[hbt!]
\centering
\includegraphics[width=0.5\textwidth]{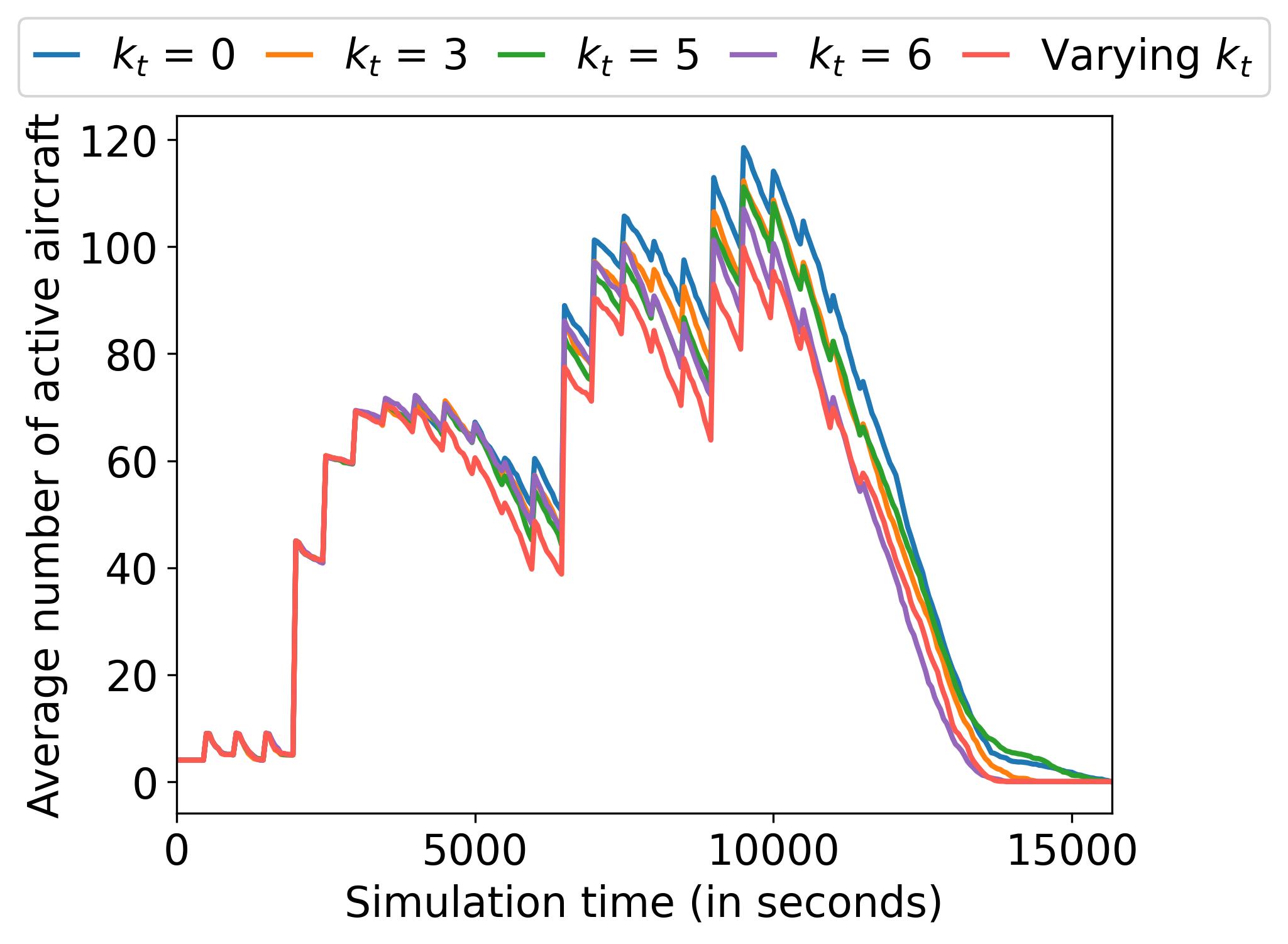}
\caption{Traffic profiles for cases when the traffic-following factor $k_t$ is fixed to specific values versus when it is varying throughout the simulation. Aircraft were introduced into the system at the same time points for all cases.} 
\label{sec1_num}
\end{figure}

Fig. \ref{sec1_num} shows how the number of aircraft in the airspace varied over the course of the simulation across different cases, even though they were introduced into the airspace at the same time in each case. Although they all showed similar traffic volumes at the beginning of the run, the case with no traffic-following, or $k_t = 0$, had the highest values, whereas the case where aircraft were varying their degree of traffic-following behavior consistently had the lowest numbers of active aircraft. This indicates that the airspace was most congested when $k_t = 0$ and least congested when $k_t$ was permitted to vary.

% \begin{figure}[hbt!]
% \centering
% % \includegraphics[width=0.5\textwidth]{total_airspace_ent.png}
% \includegraphics[width=0.5\textwidth]{sec1_num.png}
% \caption{Traffic profiles for cases when the traffic-following factor $k_t$ is fixed to specific values versus when it is varying throughout the simulation. Aircraft were introduced into the system at the same time points for all cases.} 
% \label{sec1_num}
% \end{figure}

\begin{figure}[hbt!]
\centering
\includegraphics[width=0.5\textwidth]{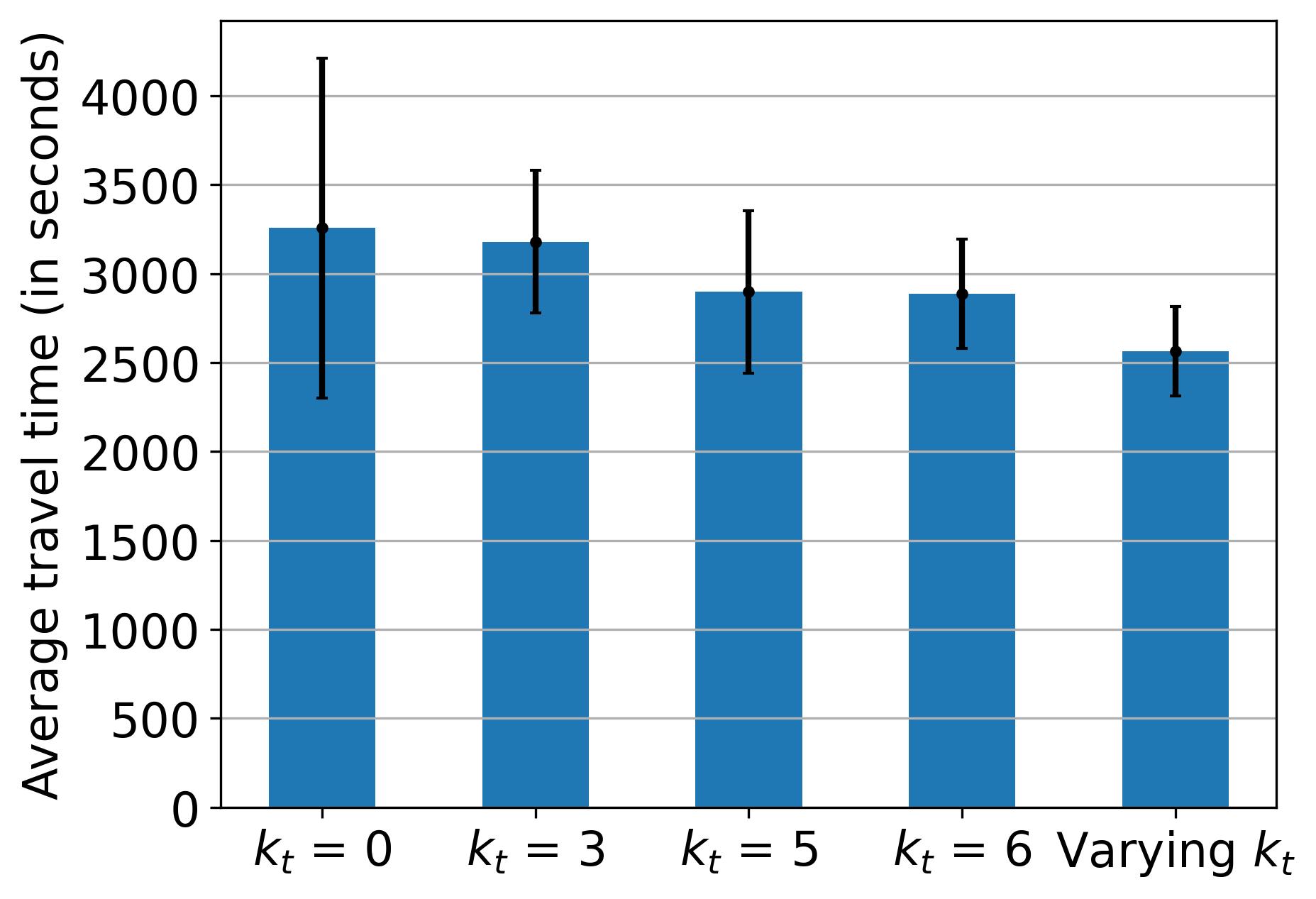}
\caption{Average aircraft travel times for cases when the traffic-following factor $k_t$ is fixed to specific values versus when it is varying throughout the simulation.} 
\label{sec1_ttd}
\end{figure}

Results for the average time it took an ownship
to traverse the airspace in each case are shown in Fig. \ref{sec1_ttd}. As expected, for cases where the traffic-following factor was fixed, increased traffic-following behavior led to lower travel times.

We observed that travel times were 11\% lower (p-value = 0.0045) when the degree of traffic-following behavior was adjusted during the course of the simulation compared to when the traffic-following factor was constant and set to its highest setting: 6. Travel times were highest when there was no traffic-following behavior. Compared to the case where there was no traffic-following at all ($k_t$ = 0), the case where aircraft adjust their degree of traffic-following behavior showed a 21\% reduction in travel time. This agrees with our past findings and intuition.

Travel times for the case where aircraft adjust their degree of traffic-following behavior should always be lower than or equal to the lowest travel time from when aircraft have fixed degrees of traffic-following behavior. This is because past results have shown that for fastest travel times, aircraft do not need to follow traffic patterns at low densities but must do so at high densities. By allowing an aircraft to adjust its degree of traffic-following behavior based on the amount of congestion, we enable it to always pick a gain that will yield the best travel time. This highlights a key benefit of aircraft independently adjusting their traffic-following behavior as needed.

\begin{figure}[hbt!]
\centering
\includegraphics[width=0.5\textwidth]{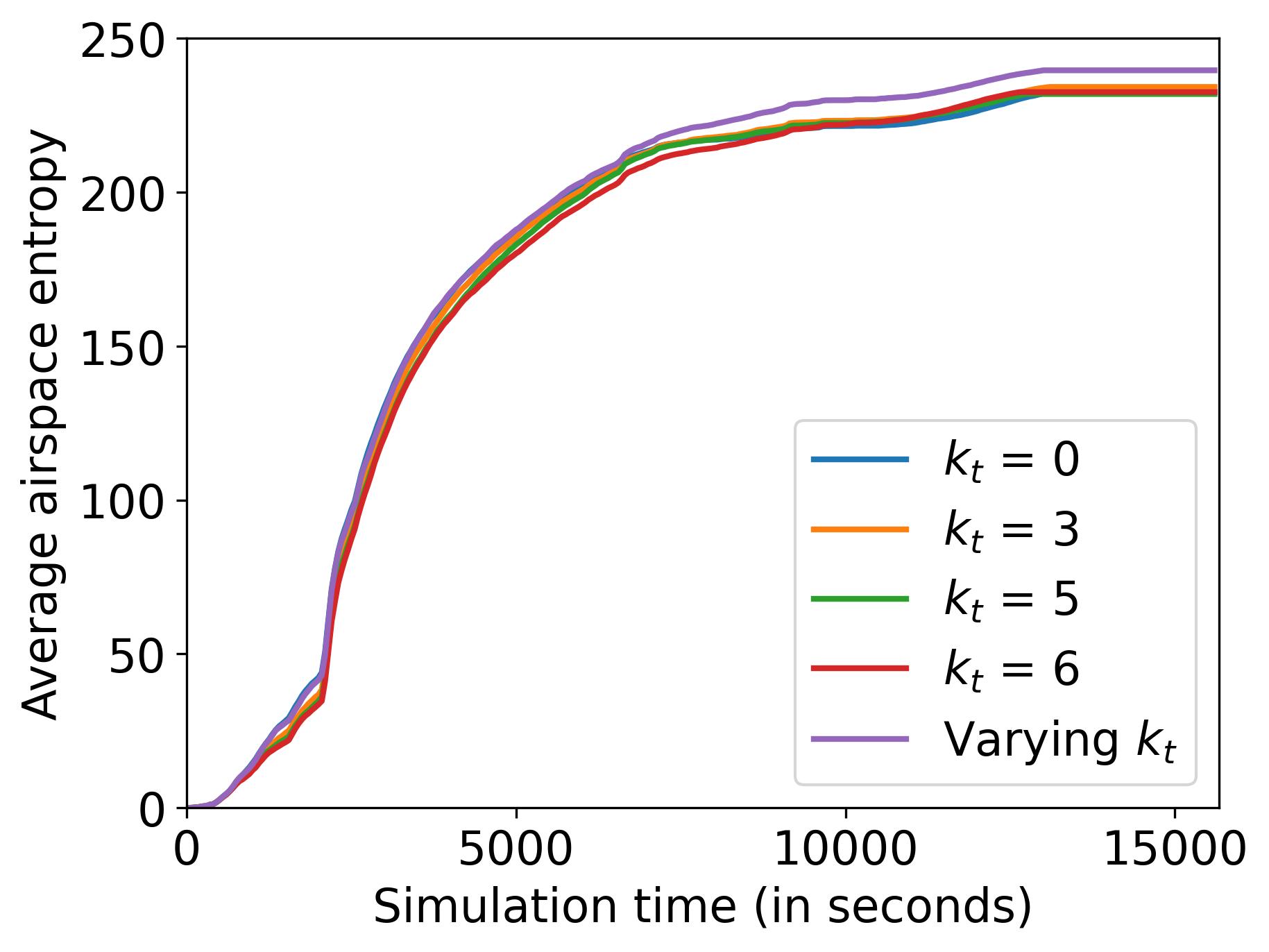}
\caption{Average airspace entropy for cases when the traffic-following factor $k_t$ is fixed to specific values versus when it is varying throughout the simulation.} 
\label{sec1_entropy}
\end{figure}

Fig. \ref{sec1_entropy} shows the average amount of airspace entropy over the course of the simulation for the different cases. As discussed in Section \ref{subsec:disc}, in each case, the total airspace entropy of a system is expected to either increase or stay constant over time, but never decrease.

It is observed that airspace entropy was lowest when there was a high degree of traffic-following behavior, i.e., $k_t$ = 6, and highest, albeit not by much, when aircraft were allowed to vary their degree of traffic-following behavior. This is because, for high $k_t$ values, aircraft chose to navigate through traversal edge pairs previously used by other aircraft, resulting in more airspace order. On the other hand, when aircraft were allowed to vary their degree of traffic-following behavior, aircraft were observed to follow traffic patterns only when beneficial. This ultimately allows aircraft to use new paths as necessary, resulting in the traversal of an overall increasing number of new pathways in the airspace, which leads to higher entropy. 

These results indicate that enabling aircraft to adjust their degree of traffic-following behavior during flight is beneficial in terms of travel time, without significant penalties in additional airspace disorder.

\subsection{The effect of varying $R_s$ on travel time and airspace entropy}\label{subsec:varying_ranges}

In this subsection, we expand on the case of aircraft varying their traffic-following factor from the last section. In particular, we study the effects of varying $R_s$ on aircraft travel times and airspace entropy. Cases with four ranges were considered: $R_s$ = \{15, 25, 35, 50\} miles. The last case covers the entire grid since the diameter of the grid is 47.697 miles. Range values were consistent for all aircraft within the same run. For example, in the simulation for $R_s$ = 15 miles, all aircraft had a range of 15 miles.

\begin{figure}[hbt!]
\centering
\includegraphics[width=0.5\textwidth]{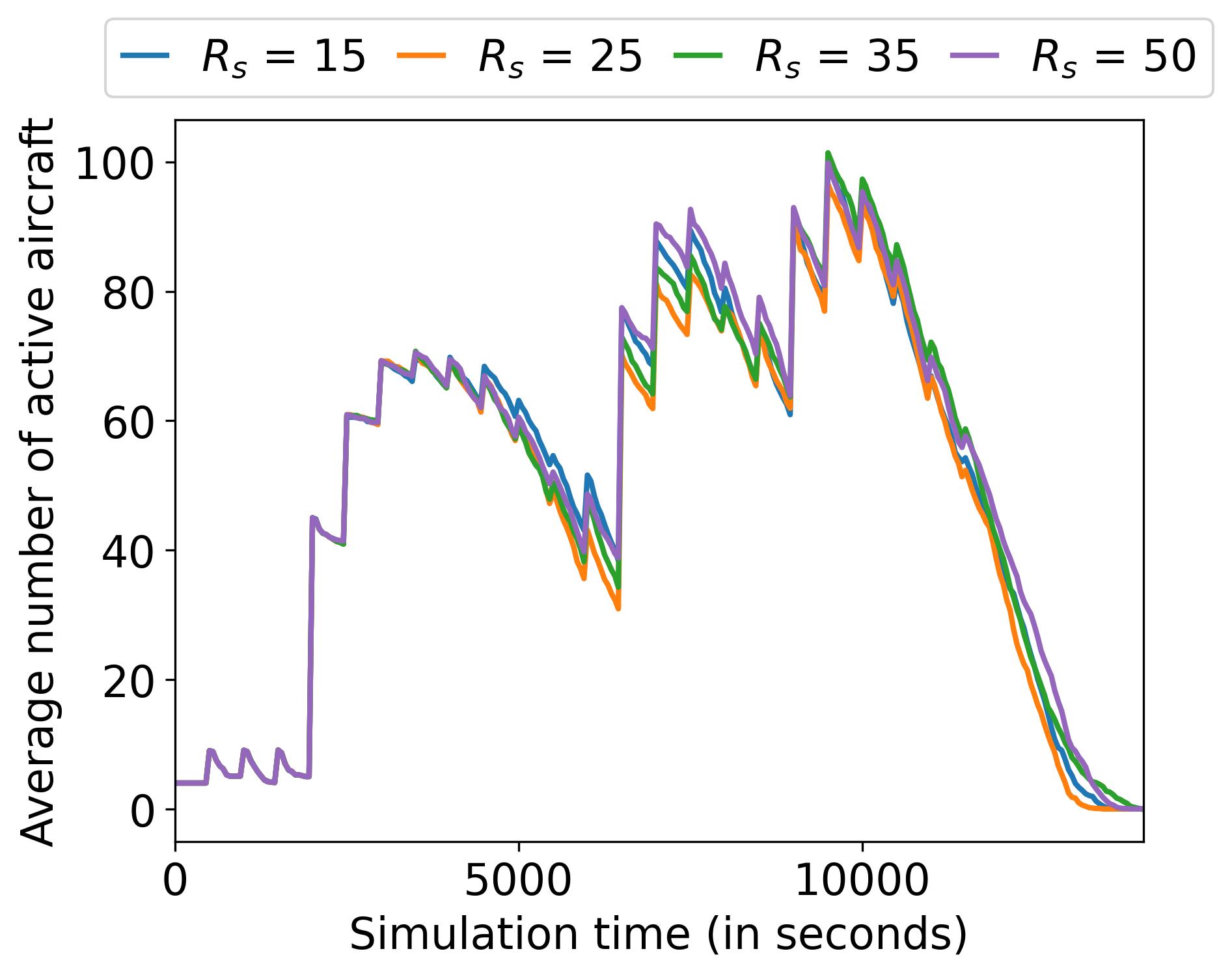}
\caption{Effect of varying the range ($R_s$) on the average number of aircraft in the airspace, with all $R_s$ values in miles. Aircraft were introduced at the same time points in all cases.} 
\label{sec2_num}
\end{figure}

Fig. \ref{sec2_num} shows how the average number of aircraft in the airspace varied over the course of the simulation across different cases, even though they were introduced into the airspace at the same time in each case. Although they all showed similar traffic volumes at the beginning of the run, cases with larger range values ($R_s= \{35 \text{ miles}, 50 \text{ miles}\}$) had the highest number of active aircraft towards the end of the simulation. The case when $R_s= 25$ miles consistently had the lowest number of aircraft in the airspace. This indicates that in cases when $R_s= \{35 \text{ miles}, 50 \text{ miles}\}$, the airspace was most congested; while it was least congested when $R_s = 25$ miles.

\begin{figure}[hbt!]
\centering
\includegraphics[width=0.5\textwidth]{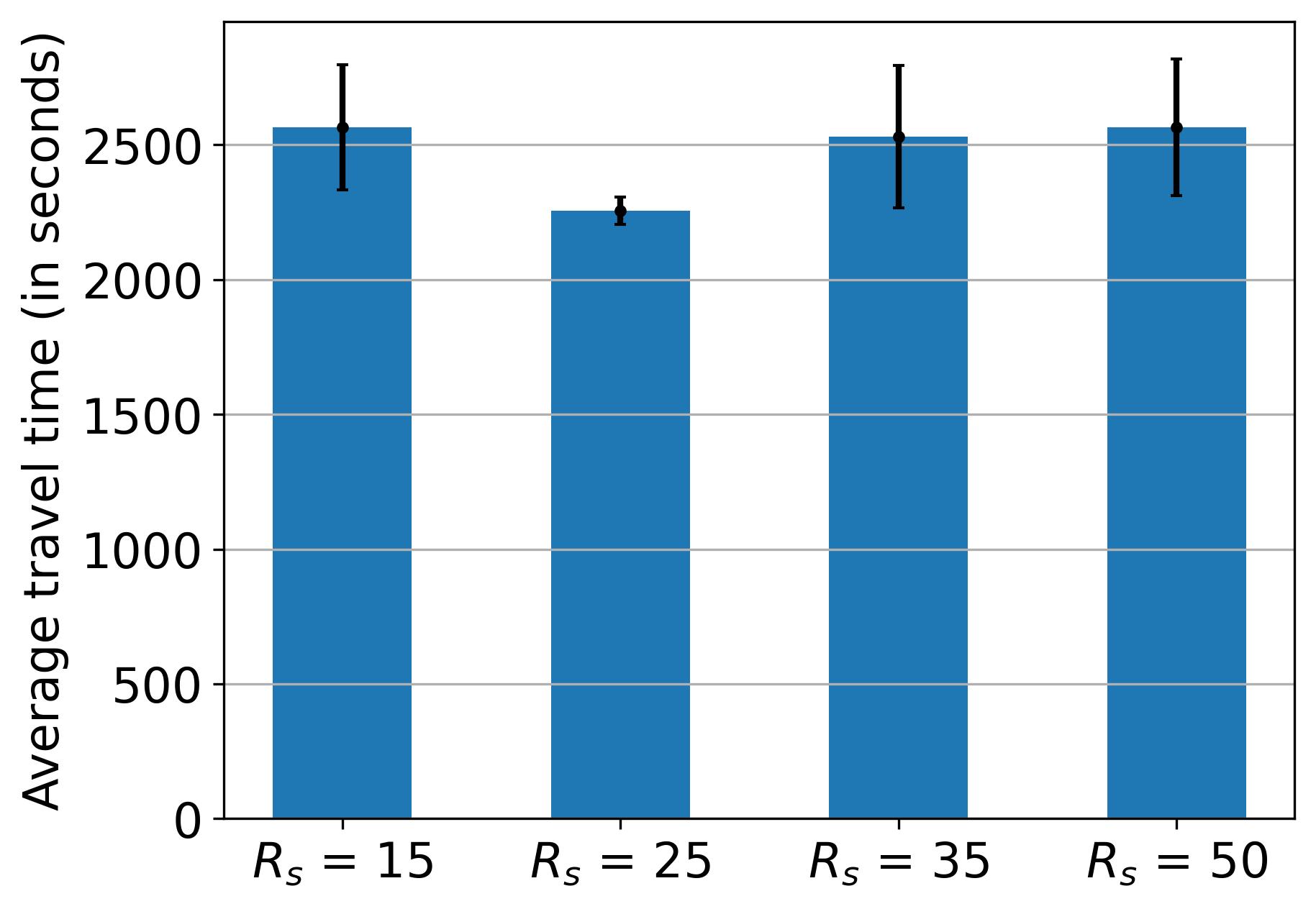}
\caption{Effect of varying the range ($R_s$) on the average aircraft travel times, with all $R_s$ values in miles.} 
\label{sec2_ttd}
\end{figure}

Results for the average amount of time it took an ownship to traverse the airspace in each case are shown in Fig. \ref{sec2_ttd}. Aircraft reached their destination the fastest when $R_s = 25$ miles, with similar travel times in all other cases.

These results agree with intuition because in the case when the range is smaller, for example $R_s = 15$ miles, aircraft might not have enough information about airspace traffic to decide whether they need to follow other traffic or not. On the other hand, when the range is too large, such as in $R_s$ = \{35 miles, 50 miles\}, an ownship might detect higher traffic densities simply because the range is larger. However, adopting a high degree of traffic-following behavior because of this might not ultimately be beneficial. For example, if one corner of the airspace is congested, an aircraft in a less congested corner might not benefit from following traffic. 

These results introduce the notion of an ideal range in this context. Under the conditions studied in this work, a range of 25 miles might be most beneficial since the case of $R_s = 25$ miles produced a 12\% reduction in travel time (p-value = 2.17E-08) compared to other ranges. This number might vary depending on several factors such as the traffic profiles studied, size of the airspace cells, and aircraft speeds.

\begin{figure}[hbt!]
\centering
\includegraphics[width=0.5\textwidth]{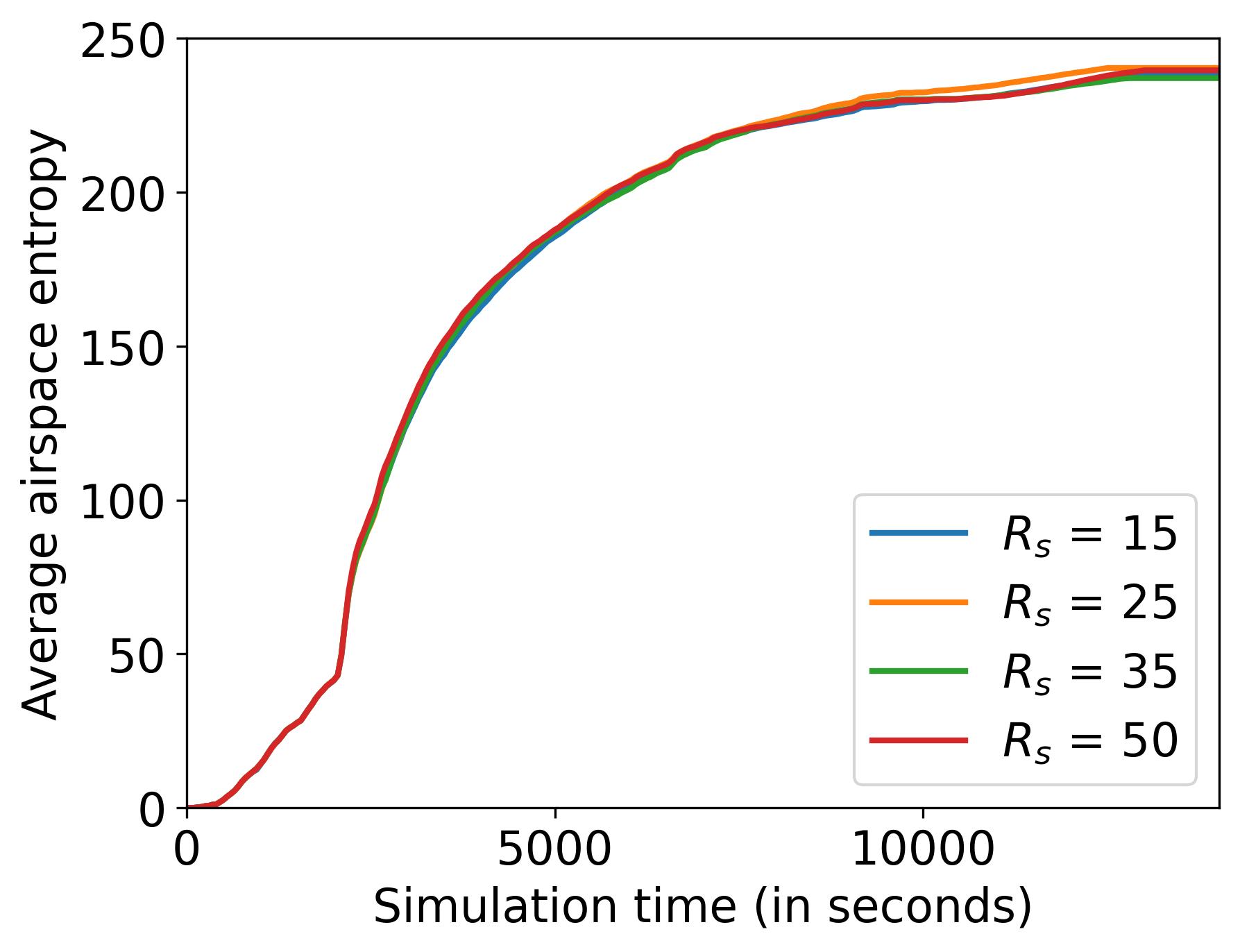}
\caption{Effect of varying the range ($R_s$) on the total airspace entropy, with all $R_s$ values in miles.} 
\label{sec2_entropy}
\end{figure}

Fig. \ref{sec2_entropy} shows the average amount of airspace entropy over the course of the simulation for the different cases. Similar to \ref{subsec:const_vs_varying}, airspace entropy remained at similar levels for all cases. This indicates that, while there might be travel time benefits that certain ranges offer, there is no significant change in the amount of disorder introduced into the airspace.

\section{Potential Applications} \label{sec:apps}
Dynamic traffic-following behavior hold significant potential for real-world applications today and in the future, spanning across both unstructured and structured airspaces. In this section, we discuss how this can be leveraged to improve safety, efficiency and scalability in diverse airspace environments.

% Dynamic traffic-following behavior can be leveraged to improve safety, efficiency and scalability in diverse airspace environments. In this section, we discuss the significant potential of this behavior for real-world applications today and in the future, spanning across both unstructured and structured airspaces. 

This work was developed to support the shift towards decentralized air traffic operations. By enabling autonomous behavior of aircraft in distributed settings, it helps create structure in airspaces that lack predefined structural procedures- such as uncontrolled airspaces or those designed specifically for higher levels of autonomy, such as UTM and U-Space frameworks, or Advanced Air Mobility environments. Additionally, this methodology can be leveraged in airspaces where existing structural guidelines are no longer useful, for example, due to weather. Simply increasing the traffic-following factor for aircraft in such airspaces will make them follow surrounding traffic, dynamically generating new pathways, while effectively updating structural procedures for future traffic.

Today, aircraft navigate to their destinations using direct route planning, maximizing airline profits. However, this can often result in airspace congestion under high traffic densities. In such scenarios, increased traffic-following behavior by individual vehicles may be used to organize airspace traffic. As shown by our results, this reduces congestion levels, increases airspace order, and lowers average travel times for aircraft. As traffic density decreases, vehicles can seamlessly return to their direct route paths by adjusting their traffic-following factor to lower levels. This dynamic adjustment of traffic-following behavior based on the state of the airspace will allow aircraft to minimize their travel times whenever possible, resulting in lower costs and  emission levels. Another option is to assign localized high-density areas predetermined high traffic-following factors. This will ensure aircraft self-organize to maintain orderly flow in high congestion hotspots. 

Aircraft often also have pre-assigned arrival windows at their destination airports that they aim to adhere to. The methodology proposed here can be used to allow aircraft to self-organize in the flow leading up to a common destination (fix or waypoint) to meet their assigned windows. If they are unable to meet these arrival time windows, aircraft can optimally self-sequence and self-schedule by further leveraging this methodology, thus reducing the need for centralized scheduling and sequencing of arrival flows. 

Traffic-following behavior can be leveraged to integrate aircraft into traffic patterns. Traffic patterns are used to predict aircraft intention, which helps lower the uncertainty of the airspace- a key factor in enabling safe air traffic operations. These patterns can be used by collision avoidance systems to help maintain separation distances among aircraft, and are also often used by controllers to lower airspace complexity and maintain workload levels under acceptable thresholds. By leveraging traffic-following behavior, aircraft can integrate into both historical and dynamically evolving traffic patterns. This can be useful while following other aircraft in instances such as platooning, following path-finding aircraft through weather systems, or while landing around airport terminals. 

Additionally, discounting the traffic pattern map over time allows aircraft greater adaptability to the current state of the airspace. For example, landing patterns around runways, often determined by current rather than older historic wind directions, can be more accurate using temporal discounting of the traffic pattern map.

When traffic patterns shift over a longer period of time, the traffic-following methodology presented here can be leveraged for airspace design. This technique enables airways to emerge in the airspace based on recent traffic patterns. For instance, if a geographical region's demand changes over time, we might see changes in the amount of air traffic around that area. The methodology presented in this paper can be used to redesign the airspace's structural procedures to improve efficiency.

Overall, this methodology fosters order among autonomous agents, enabling seamless interaction and integration within the broader system. This technique can be used for creating order in any multi-agent system, from airspace traffic management to robot swarms used in delivery fulfillment warehouses.

% This technique can optimize multi-agent coordination across various systems, from airspace management to delivery fulfillment warehouses. 

% Dynamic traffic-following behavior holds significant potential for real-world applications. For example, direct route planning by operators can often cause airspace congestion under high traffic densities. Dynamic traffic-following can reduce congestion, improve order, and lower travel times in dense airspaces, while allowing a return to direct routes when traffic decreases. 

% Traffic-following behavior can be leveraged to integrate aircraft into traffic patterns, for instance around airport terminals or in helping aircraft follow other aircraft with dynamic paths through weather systems. 

% Additionally, discounting the traffic pattern map over time allows aircraft greater adaptability to the current state of the airspace. For example, landing patterns around runways, often determined by the current rather than older historic wind directions, can be more accurate using temporal discounting of the traffic pattern map. Overall, this methodology fosters order among autonomous agents, enabling seamless interaction and integration within the broader system. This technique can optimize multi-agent coordination across various systems, from airspace management to delivery fulfillment warehouses. 

\section{Conclusion} \label{sec:future}
In this paper, we studied the emergence of orderly traffic in a distributed, autonomous multi-agent system. Past studies showed that under high density conditions, order generated from traffic-following behavior is beneficial for travel times, whereas under low densities, choosing direct paths is more beneficial. In this paper, we extended our methodology to allow aircraft to dynamically adjust their degree of traffic-following behavior during flight as a function of the density of traffic within a spatial range. This allowed aircraft to follow other traffic only when beneficial, while always resulting in the best travel times. Additionally, we explored the impact of continuously discounting past traffic pattern information over time to ensure more recent patterns are followed when they differ from historic ones. This ensures aircraft adapt to the current state of the airspace, which further reduces aircraft travel times.

Sensitivity analyses revealed that enabling an aircraft to adjust its traffic-following factor during flight, as opposed to maintaining it at fixed levels, improved travel times at the cost of minimal levels of additional airspace disorder. Furthermore, we identified the existence of an optimal spatial range within which an aircraft should follow traffic.

Overall, the study presented here is an important extension. By having a technique to dynamically organize traffic when necessary, we are able to maximize the throughput of an airspace. By extending the current study, in the next phase we will determine the limit at which creating order within the airspace is no longer feasible or beneficial. Once this limit is reached, we can investigate flow management practices to ensure safety of the airspace while enabling the best use of airspace resources, thus closing the loop on self-limitation behavior in the airspace. For this, we will introduce metrics such as aircraft trajectory flexibility to measure safety and improve existing models to incorporate additional degrees of freedom. Incorporating different altitudes, introducing vehicle heterogeneity, and applying this methodology to real traffic data are also important directions for future studies.

% Incorporating altitude and then comparing four-dimensional trajectories generated from this scheme to direct planning trajectories generated by airlines are also important extensions.

Ultimately, we have demonstrated that incorporating self-organizing, orderly behavior is beneficial and potentially critical, particularly in high-density airspaces, to enable scalable concepts that afford higher levels of autonomy to both operators and vehicles. In the broader context, this work has encouraged us to consciously examine desired behaviors and metrics in the design of multi-agent systems.

% Discounting the traffic pattern map with time is useful in the context of winds and weather. For example, landing patterns around runways, often determined by the current rather than older historic wind directions, can be more accurate using temporal discounting of the traffic pattern map. 
% Aircraft can use this methodology to follow other aircraft while finding paths through weather systems.

% For instance, an aircraft navigating through airspace during a weather phenomenon can better understand how other aircraft are currently maneuvering, enabling it to make more informed decisions about its own trajectory.

\section*{Acknowledgment}

This work was funded by the NASA Convergent Aeronautics Solutions and Transformative Tools and Technologies projects. Special thank you to David Wing and Andy Lacher.

\printbibliography

\section*{Biographies}

\textbf{Anahita Jain} is a Ph.D. candidate in the Aerospace Engineering and Engineering Mechanics Department at The University of Texas at Austin. Her research interests lie at the intersection of optimization and algorithm design of nature-inspired multi-agent systems.

\textbf{Dr. Husni Idris} is an aerospace research engineer at the NASA Ames Research Center, where he leads research in collective autonomous mobility for enabling increasingly autonomous airspace operations.
% He holds a Ph.D. from MIT and is an associate fellow of AIAA.

\textbf{Dr. John-Paul Clarke} is a professor of Aerospace Engineering and Engineering Mechanics at The University of Texas at Austin, where he holds the Ernest Cockrell Jr. Memorial Chair in Engineering. 

\textbf{Dr. Daniel Delahaye} is the head of the optimization and machine learning team at ENAC, where he conducts research on stochastic optimization for airspace design and large-scale traffic assignment.

% \aj{Yutian's comments:
% \begin{enumerate}
%     \item for a journal, you can have pictures from diff timestamps with diff rows showing the diff cases
%     \item try log for entropy figures?\
%     \item can have a much more detailed intro and lit review section
%     \item another section for future work and applications
% \end{enumerate}}

% \bibliography{sample}
\end{document}